 \documentclass[10pt,a4paper]{article}
\usepackage{epsfig}
\title{ Cylindrical Symmetry Discrimination of \\Magnetoelectric Optical Systematic Effects in a Pump-probe Atomic
Parity Violation Experiment }  
\author{  M-A.~Bouchiat, J. Gu\'ena, M. Lintz    \vspace{5mm} \\
     {\small         Laboratoire Kastler Brossel\footnote{Laboratoire de l'Ecole
Normale Sup\'erieure associ\'e au CNRS (UMR 8552) et \`a l'Universit\'e Pierre et Marie
Curie}~~et F\'ed\'eration de Recherche\footnote{F\'ed\'eration de Recherche de l'Ecole Normale Sup\'erieure 
                     associ\'ee au CNRS (FR684)}}\\
{\small D\'epartement de
                Physique de l'Ecole Normale Sup\'erieure,}\\ 
 {\small                24 Rue
               Lhomond, F-75231  Paris  Cedex 05, France} \\}
  
\begin{document}
  \maketitle
 \bibliographystyle{unsrt} 
 \begin{abstract}
A pump-probe atomic parity violation (APV) experiment performed in a longitudinal electric field $\vec E_l$, has the
advantage of providing a signal which breaks mirror symmetry but preserves cylindrical symmetry of the set-up, {\it i.e.}
this signal remains invariant when the pump and probe linear polarizations are simultaneously rotated about their common
direction of propagation. The excited vapor acts on the probe beam as a linear dichroic amplifier, imprinting a very specific
signature on the detected signal. Our differential polarimeter is oriented to yield a null result unless a chirality of some kind
is acting on the excited atoms. Ideally, only the APV ($\vec E_l$-odd) and the calibration ($\vec E_l$-even) signals should
participate in such a chiral atomic response, a situation highly favourable to sensitive detection of a tiny effect.  In the present
work we give a thorough analysis of possible undesirable defects such as spurious transverse fields or misalignments, which
may spoil the ideal configuration and generate a chiral response leading to possible systematics. We study a possible way to
get rid of such defects by performing global rotations of the experiment by incremental angular steps $\phi$, leaving both
stray fields and misalignments unaltered. Our analysis shows that at least two defects are necessary for the
$\vec E_l$-odd polarimeter output to be affected; a $\cos{(2\phi)}$ modulation in the global rotations reveals the
transverse nature of the defects. The harmful systematic effects are those which subsist after we average over four
configurations  obtained by successive rotations of 45$^\circ$. They require the presence of a stray transverse electric field.  
 By doing auxiliary atomic measurements made in known, applied, magnetic fields which amplify the systematic effect, it is
possible to measure the transverse E-field and to minimize it.   Transverse magnetic fields must also be carefully compensated
following a similar procedure.  We discuss the feasibility of reducing the systematic uncertainty below the one percent level. We
also propose statistical correlation tests  as  diagnoses of the aforementioned systematic effects.   

  ~~~~

\noindent  PACS. 32.80.Ys - 32.60.+i - 33.55.Fi - 42.25.Lc               

\end{abstract}
\section{Introduction}
Atomic Parity violation (APV) experiments have been motivated by their ability to probe neutral current weak interactions in conditions very
different from particle physics experiments of all kinds and hence to yield valuable complementary
information 
\cite{bou97}. They probe the electron-quark electroweak interaction at distance scales very different from
those explored in high energy measurements. Moreover, in atoms all the quarks contribute coherently while at high
energies the nuclei and even the nucleons are broken and the quarks act independently. In atoms the issue at stake is the
detection of a tiny electric dipole transition amplitude, strictly forbidden by the laws of electromagnetism, but 
allowed by the weak interaction which breaks mirror-symmetry.  APV  can ``show up'' in several ways:
optical rotation in allowed M$_1$ transitions
\cite{Bi91,Pb93,Tl95}, or electroweak interference effects in the transition probability of an 
$E$-field-assisted forbidden atomic transition \cite{bou82}, which affects either the population
\cite{dre85,gil86,woo97,ben99,woo99} or the orientation in the upper state \cite{bou82}. In the latter case, the detection always
relies on some fluorescence light monitoring with or without polarization analysis. Since the interference term involves the
product of the parity violating amplitude
$E_1^{pv}$ and the $E$-field induced amplitude,
$\beta E$, while the transition rate is proportional to $\vert \beta E \vert^2$, the left-right asymmetry, $\propto {\rm
Im}~E_1^{pv}/
\beta E $, decreases as 
$1/E$. More recently, our group has demonstrated a novel kind of pump-probe experiment \cite{bou02}. Here the
6S-7S highly forbidden transition is excited in a longitudinal electric field $\vec E_l$ by an intense  pulse of resonant light
which lasts for a time shorter than 
the 7S lifetime. It is immediately followed by the light pulse of a second beam, the
probe, resonant with the 7S-6P$_{3/2}$ allowed transition and colinear with the excitation beam. For short pulse durations,
the population inversion produced by the pump is sufficient to produce transient amplification of the probe beam. APV
 shows up because the probe amplification depends on the relative orientation of the linear 
polarization of the  excitation laser $\hat
\epsilon_{ex}$  and the probe laser 
$\hat \epsilon_{pr}$. Specifically, there is a chiral contribution to the optical gain of the vapor characterized by
the pseudoscalar
$(\hat \epsilon_{ex} \cdot \hat
\epsilon_{pr})(\hat \epsilon_{ex} \wedge \hat \epsilon_{pr} \cdot \vec E_l)$ which takes opposite values for two
mirror-image configurations, as for instance those observed in the two channels of our polarimeter monitoring $\hat
\epsilon_{pr}$ (\S~3). A 9$\%$ accurate measurement
\cite{bou02}, being currently improved, has validated  the method. One of its advantages is to provide an
independent method of APV measurement. It is well known that one of the main difficulties in measuring the very
small APV effects lies in the discrimination against systematic effects. These have different origins depending on
the chosen configuration, hence the importance of a new configuration. In addition, this stimulated emission detection
scheme benefits from several attractive features: dark-field detection of the left-right asymmetry, reliable, line shape
independent calibration procedure, amplification of the asymmetry itself.  For instance, the right-left asymmetry instead of
being a decreasing function of the applied
$\vec E_l$-field benefits from an amplification mechanism via propagation of the probe beam through the optically thick
excited vapor \cite{cha98}. 

Moreover, the cylindrical symmetry of the experiment, another original feature, plays an important role: the
amplification asymmetry is expected to remain invariant under simultaneous rotations of the polarizations $\hat
\epsilon_{ex}$ and $\hat \epsilon_{pr}$  around their axis of propagation \cite{gue98}. In the present paper, we explain
how  this property can be exploited to discriminate against parity conserving (PC) signals generated by imperfections, 
because of their variation under these rotations. Although some of these signals simulate the PV effect in a
given polarization configuration, their signatures are signals which break cylindrical symmetry.  A schematic
of the ideal experiment represented on Fig. 1 shows the two orthogonal symmetry planes defined by the electric field $\vec
E_l$ and the linear excitation polarization
$\hat
\epsilon_{ex}$. APV gives rise to a tilt of the optical axes of the excited vapor out of those planes. The incoming probe
polarization chosen either parallel or perpendicular to that of the excitation beam 
provides a superposition of the two configurations of opposite handedness,  $\hat \epsilon_{ex},
\hat
\epsilon_{pr}^{X},
\vec E_l$, and $\hat \epsilon_{ex}, \hat \epsilon_{pr}^{Y},
\vec E_l$, analyzed simultaneously by our polarimeter ($\hat \epsilon_{pr}^{X}$ and $\hat \epsilon_{pr}^{Y}$ 
denoting the two components of the probe  polarization at
$+$ and $- 45^{\circ}$ to its input direction). Since our polarimeter operates in a balanced mode, the probe amplification
difference, {\it i.e.} the right-left asymmetry, is directly extracted for each excitation laser pulse from the optical signals
$S_1, S_2,$ recorded for the two channels. {\it The
PV left-right asymmetry is expected to remain invariant under a global rotation of the experiment, performed by rotating
the polarimeter and beam polarizations altogether around the common beam axis.} This important test can reveal defects
such as transverse $\vec E$ or $\vec B$ fields which  
remain fixed in the laboratory frame while the polarizations are rotated. We discuss how this can be used to
minimize  the systematic errors below a known level. This is even more important given that we are currently improving
our experiment with the aim of reducing the statistical error down to the one percent level.    
\begin{figure}
\centerline{\hspace{-15mm}\epsfxsize=150mm \epsfbox{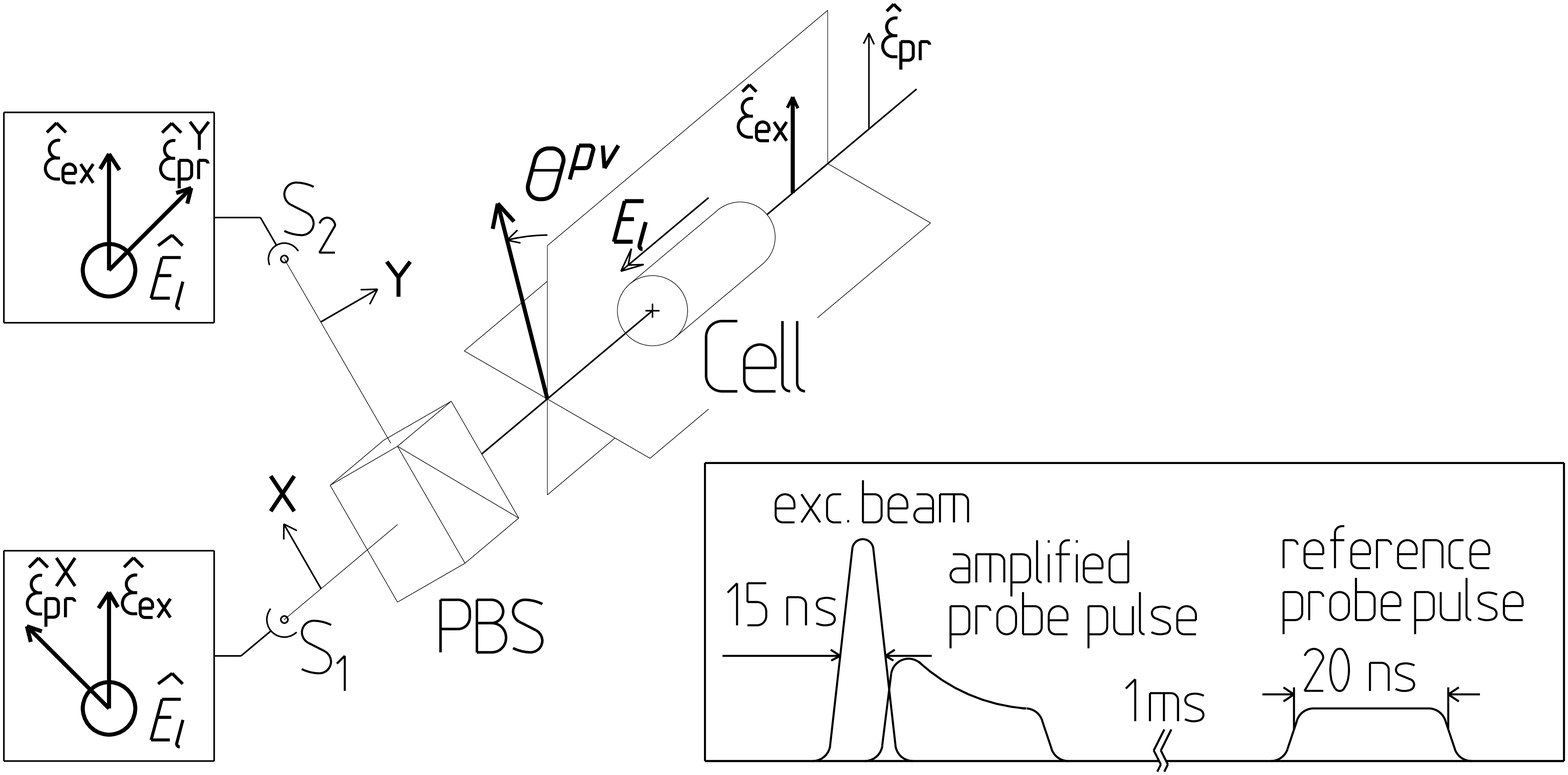}}
\vspace{-25mm}
\caption {\footnotesize Schematic of the experiment showing the two orthogonal symmetry planes defined 
by the electric field $\vec E_l$ and
the linear excitation polarization  $\hat
\epsilon_{ex}$. APV gives rise to a tilt
$\theta^{pv}$  of the optical axes of the excited vapor out of these planes. The incoming probe polarization $\hat
\epsilon_{pr}$ provides a  superposition of the left and right-handed ($\hat \epsilon_{ex}, \hat \epsilon_{pr}^X, \vec E_l$
and $\hat \epsilon_{ex}, \hat \epsilon_{pr}^Y, \vec E_l$) 
configurations   analyzed. The probe amplification difference is 
directly extracted from the optical signals S$_1$, S$_2$, recorded in each channel of the Polarizing Beam Splitter 
(PBS). Inset: timing of the experiment repeated at 100 Hz.}
\end{figure}

 In this paper we first explain the origin of the chiral optical gain in terms of a parity violating contribution to the atomic alignment
in the 7S state induced by linearly polarized 6S-7S excitation (\S~2). Next, we introduce a general formalism necessary for a
rigorous treatment of experimental defects (\S. 3). The basic principle of our polarimetry measurements is presented and
used to obtain the relation between the PV atomic alignment in the excited state and the ``atomic'' imbalance of the
polarimeter (\S. 3).  This relation is useful to predict the variations of the signals generated by defects  under simultaneous
rotations $ R(\hat k ,  
\phi)$ of the input polarizations $\hat \epsilon_{pr}$ and 
$\hat \epsilon_{ex}$ about the common beam direction $\hat k$ by an angle $\phi$, which leave invariant the APV and
the calibration signals.  As an example we treat the perturbation induced by the simultaneous presence of transverse
electric and magnetic fields, 
$\vec E_t$ and $\vec B_t$. We show that this leads to a parity conserving, magnetoelectric optical effect,
 which,  depending on the property of $\vec E_t$ and $\vec B_t$ under the $\vec E_l$-field reversal, may
simulate the PV signal in a fixed $\hat \epsilon_{ex}$, $\hat \epsilon_{pr}$ configuration, but which exhibits characteristic
modulations under rotations of the polarizations (\S.~4). In this particular case, it is easy to arrive at a
judicious choice of the rotation angle $\phi$. We show that only four different $\hat
\epsilon_{ex}$, $\hat
\epsilon_{pr}$ configurations are necessary to extract maximal information about the systematic effect generated by the stray fields: 
namely those generated from an initial configuration by three steps of successive 45$^{\circ}$ rotations of both $\hat
\epsilon_{ex}$ and 
$\hat \epsilon_{pr}$.  This result can be extended to systematics arising from two transverse magnetic
fields, one odd, and the other even in
$\vec E_l$-reversal (\S. 5). This property is further extended and can be associated to the most general structure of the
atomic density matrix for the 7S state. We show that a misalignment of the excitation and probe beams can
have the same effect as a transverse magnetic field. Finally, taking into account the possible magnitude of the residual defects
on our set-up, we discuss  at what level we might reasonably expect to reduce the systematic uncertainty 
(\S~6.). In addition, we present a quite independent diagnosis based on a statistical test to be performed on the PV data which
could reveal the presence of a harmful systematic effect of magnetoelectric origin.
\section{ APV manifestation via a chiral optical gain}
\subsection{APV contribution to the angular anisotropy in the excited state}
When an atomic vapor is excited with a {\it linearly} polarized laser beam, resonant at the frequency of an $E_1$ allowed
transition, it is easily verified that the excited state posseses an {\it alignment} with a privileged direction determined by the
polarization of  the excitation beam. More
precisely the quantum average of the operator 
$\vert\vec F\cdot
\hat \epsilon_{ex} \vert ^2 $ taken over the density matrix $\rho_e(t=0)$ of the 7S,F' state at the excitation time $t=0$, differs
from the typical value for  an isotropic distribution, {\it i.e.} ${\rm Tr} \{\rho_e(0)  \vert\vec F\cdot
\hat \epsilon_{ex} \vert ^2 \} \not = \vert \vec F  \vert ^2 /3 \cdot {\rm Tr} \{\rho_e(0) \}$.   
 
Here we excite the vapor via one hyperfine component  $6S,F-7S,F'= F\pm 1 $ of the
highly forbidden transition. The effective transition dipole
$\vec d^{eff} $ can be written \cite{bou97,bou75}:
\begin{equation}
\vec d^{eff}  = - i \beta \vec \sigma \wedge \vec  E_l + M_1 \vec \sigma \wedge \hat k  - i {\rm Im}E_1^{pv} \vec \sigma \; .
\end{equation}
The components of the electronic spin operator $\vec
\sigma$ are the Pauli spin matrices. The first contribution is the parity conserving (PC) amplitude induced by the applied
electric field parallel to the wave vector
$\hat k \parallel
\hat z$ of the excitation beam and associated with the vector part of the transition polarizability $\beta$.  The scalar 
part $\vec d = - \alpha \vec E$ can be ignored on a $F \rightarrow F' = F\pm 1$
transition. Hyperfine mixing in presence of  a transverse magnetic field will be considered in \S ~4.5. The second
contribution is associated with the
$M_1$ amplitude, but it will be shown later on (\S~2.4) to lead only to very small effects. The last term in Eq. 1 arises
from the PV electric dipole amplitude $E_1^{pv}$ characteristic of the weak interaction.  If we ignore for the moment the
$M_1$ contribution and if we choose  $\vec E_l =\pm  E_l \hat z$, we can rewrite the transition amplitude 
as follows:
\begin{equation}
\vec d^{eff}\cdot \hat \epsilon_{ex} = \mp  i \beta E_l \; \vec \sigma \wedge \hat z \cdot ( \hat \epsilon_{ex} \pm 
\theta^{pv} \hat z \wedge
\hat \epsilon_{ex})\; , \hspace {5mm} {\rm with} \hspace{5mm} \theta^{pv}= - {\rm Im} E_1^{pv}/ \beta E_l\; . 
\end{equation} 
This equation has a simple physical interpretation: the effect of the PV transition amplitude is equivalent to a
rotation of the linear polarization $\hat \epsilon_{ex}$ about
$\vec E_l$ by the small angle
$\theta^{pv}$, typically $\sim 10^{-6}$ rad for $E_l \approx 1.7$ kV/cm. The sense of this rotation changes when the
direction of
$\vec E_l
$ is reversed. The angle 
$\theta^{pv}$  is the important parameter to be determined since it yields the ratio between the PV and the Stark amplitudes.
Provided the $E_l$-field magnitude is known, one can use a value of $\theta^{pv}$ to obtain the weak charge $Q_W$ of the
Cs nucleus by relying on Atomic Physics calculations of $E_1^{pv} / Q_W$ 
\cite{blu90,der01,mil01,fla02}. 
 \subsection{Alignment  tensor of the excited state. }
The Stark-induced transition creates  an  excited state $ 7 S,\,F^{\prime} $ endowed with an alignment. For the sake 
of completeness we summarize here some basic definitions. First we introduce the second-rank tensor  operator $
\hat{T}^{(2)}
$ constructed from the total  angular momentum $ \vec{ F} $. In order to  avoid unnecessary
algebraic  complications we shall  use a cartesian basis set:
\begin {equation}
\hat{T}^{(2)}_{i j}= \frac{1}{2}( F_i \,F_j+F_j \,F_i)-\frac{1}{3} \vec{ F}^{2} \delta_{i,j}\;. 
\label{r2tensorop}
\end{equation}

We define now  the traceless symmetric alignment tensor    $ {\cal{ A}}_{i  j} $  by  the following quantum averages :
\begin {equation}
{\cal{ A}}_{i  j}(t) ={\rm Tr} \left \{\rho(t) \, \hat{T}^{(2)}_{i j} \right \} \;, \hspace{5mm} {\rm implying}\hspace{5mm} 
\sum_{k} {\cal{ A}}_{k k}(t) =0\; .
\end{equation}
  It is convenient to introduce as a  visual tool the  alignment ellipsoid defined by the quadratic
 equation : 
$$      \sum_{i,j}  x_i \, x_j \, ( {\cal{ A}}_{i  j}+ \frac{1}{3} F^{\,\prime} (F^{\,\prime}+1)\, \delta_{i,j} )= \frac{1}{3}
F^{\,\prime} (F^{\,\prime}+1)\; . $$
In absence of  alignment  (${\cal{ A}}_{i  j}=0 $)  the  ellipsoid reduces to the unit sphere.

\subsubsection{General expression}
 For the Stark-induced transition  induced by the vector polarizability, in contrast to allowed electric dipole ones, the
preferred direction of the excitation process is not 
${\hat {\epsilon} }_{ex}$ but rather the orthogonal direction $\hat {E}_l \wedge {\hat {\epsilon}}_{ex} $. If we ignore the
PV contribution, the alignment created in the excited state can be represented by an ellipsoid 
of revolution symmetry about a direction parallel to
$\hat k \wedge \hat \epsilon_{ex}$ which preserves the planes of symmetry of the experiment
 defined by the common direction of $\vec E_l$ and $\hat k$ and by 
$\hat \epsilon_{ex}$ (see Fig. 1). The effect of the PV contribution is to tilt this ellipsoid about
$\vec E_l$ by the small angle $\theta^{pv}$. As a result, the alignment no longer preserves the symmetry planes of the
experiment: this is the manifestation of parity violation in the pump-probe experiment discussed presently.

We are going to present a formalism which  can be applied  to  more general situations  than the  Stark-induced  
 $ 6S \rightarrow 7S $ transition and which will allow us to incorporate  the  $E_1^{pv} $ and $ M_1$
 contributions  as well as  those associated with experimental  defects  breaking  the cylindrical symmetry of the 
{\it ideal experiment}. 
The effective transition operator $T_{eff} $ is then given by: 
\begin{eqnarray}
T_{eff} &=& \vec{b}\cdot\vec {\sigma},  \nonumber\\
{\rm where} \hspace {3mm} \vec{b}&=& i \,\beta  \vec{E} \wedge \hat {\epsilon}_{ex} + i\, {\rm Im}\, E_1^{pv}\,\hat
{\epsilon}_{ex} -  M_1 \hat{k}\wedge \hat {\epsilon}_{ex} \;.
\label{Teff}
\end{eqnarray} 
The direction  of $ \vec{E} $ is for the moment arbitrary. The laser selects only one hfs
component  $nS, F \rightarrow (n+1)S, F'$. The excited state density matrix, up to a normalization factor, is then given by: 
\begin{equation}
{\rho}_{e} = P_{_{F'}}\,  T_{eff} P_{_{F}} \,  \rho_g P_{_{F }} \, T_{eff }^{\dagger} \, P_{_{F'}}\,,
\end{equation}
where  $\rho_g$ is the restriction of the density operator to the $nS$ ground state. $P_{_{F}}$ is the
projector on the $nS, F$ sublevel and $P_{F'}$ the projector on the $(n+1)S, F'$ sublevel. 
Since we are mostly interested in the $6S, F \rightarrow 7S,F'$ transition with $F' \not= F$, it is convenient to
write
$\rho_g = P_{_F}$ and $ P_{_{F'}}= 1 \hspace{-0.8mm}{\rm I} - P_{_F }$. We apply the Wigner-Eckart theorem to
the spin operator $\vec \sigma$ acting in the hyperfine subspace $F$ :  
\begin{equation}
P_{_F} \,  \vec {\sigma}  P_{ _F } =  2 g_{_F} \, P_{_F} \,  \vec {F} , \hspace{10mm} {\rm where}  \hspace{5mm} g_{_F} =-g_{_{F'}}  =
2(F-I)/(2I+1), 
\end{equation}
and $I $ is the nuclear spin, in our case equal to 7/2 for natural cesium, $^{133}$Cs.
 
Using the following  identities involving the {\it c}-number vector $ \vec{b} $ :
$$ 
  P_{_{F'}}  \vec{\sigma}\cdot \vec{b} P_{_{F}} \vec{\sigma}\cdot \vec{b}^*\, P_{_{F}'}=
   P_{_{F'}} \vec{\sigma}\cdot \vec{b}\,( 1  - P_{_{F'} } )  \,\vec{\sigma}\cdot \vec{b}^*
   P_{_{F'}}=  P_{_{F'}}\,\vert\vec{b }\vert^2  -  
P_{_{F'}} \vec{\sigma} \cdot \vec{b}\,  P_{_{F'}}  \vec{\sigma}\cdot \vec{b}^*\, P_{_{F'}} , 
 $$ 
 we readily  obtain the excited state density matrix $ {\rho}_{e}(0)$ at the  excitation time $ t=0$:
\begin{equation}
 {\rho}_{e}(0)=   \left ( \vert\vec{b }\vert^2  - 4 g_{_{F'}}^2  \,(\vec{F}\cdot\vec{b} )(\vec{F}\cdot\vec{b}^* )
\right)   P_{_{F'}}\; .
\label{roe}
\end{equation}  

Later on, it will be of interest to consider  the transformation of the excited state density matrix
 under a space rotation $R$: $ {\rho}_{e}(0) \rightarrow  U(R)\,  {\rho}_{e}(0) \,U(R) ^{ \dagger}$,
 where  $U(R)$ is  the  unitary operator associated to $R$.  Using the following basic relations 
which result from the very definition of $U(R)$, we can write
$ U(R)\,  \vec{b}\cdot\vec{F} \,U(R) ^{ \dagger}= (R^{-1} \, \vec{F})\cdot \vec{b}=\vec{F}\cdot( R\,\vec{b}), $
we arrive to a simple   rule for the density matrix rotation transformation :
on the r.h.s of eq.(\ref{roe}) replace the vectors $\vec{b} $ and $\vec{b}^* $ by the rotated vectors: 
$ \vec{b}\rightarrow R\,  \vec{b}\, ,\,\vec{b}^* \rightarrow R\,  \vec{b}^*.$ 

 Let us now evaluate the  $ 7S $ alignment tensor components   ${\cal{ A}}_{i\,j}^{e}(0 ) $:
\begin{eqnarray}
 {\cal{ A} }_{i\,j}^{e}(0)&=&{\rm Tr} \{\rho_{e}(0)\,(F_i  F_j-\frac{F'(F'+1)}{3}\,{\delta}_{i,j} ) \} \nonumber \\
& = &- 4 g_{_{F'}}^2   
    \left ( {\rm Tr} \{ (\vec{F}\cdot\vec{b} )(\vec{F}\cdot\vec{b}^*) \;  F_i \, F_j   P_{_{F'}} \}  
 -   \vert\vec{b }\vert^2 (2\, F'+1)(\frac{\,F'^ \,(F'+1)}{3})^2
\,{\delta}_{i,j} \, \right ) \;.
\end{eqnarray}

In order to proceed, we  have to extract  from the rank-four  tensor operator 
$    T_{i\,j\,k\,l }=  F_{i}\, F_{j }\,F_{k}\, F_{l} $ 
the scalar pieces which are the only ones having   non-zero  traces. They are obtained by contracting two pairs
among  the four indices in all  possible ways. We arrive  in this manner  at the following expression for the needed trace, 
where we have used   its  invariance under  circular permutation  of the indices :
\begin{equation}
{\rm Tr}\{  T_{i\,j\,k\,l } \}= A ( \delta_{i\, ,j} \, \delta_{k\,,l}+ \delta_{i\,,l} \, \delta_{j\,,k})+B \,\delta_{i\, ,k} \,
\delta_{j\,,l}.
\label{trT}
\end{equation}
The rational numbers $A$ and $B$ can be easily obtained  as  linear combinations of   $ S_2 ( F') $ and $ S_4 ( F') $
 where $S_n (N) = \sum_{m=0}^{m=N}\,m^n$ 
by calculating directly $ {\rm Tr} F_z^2 $ and $ {\rm Tr} F_z^4$
and comparing with the results obtained by using eq.(\ref{trT}):
   \begin{eqnarray}
 A &= & F'(F'+1)\,  S_2 ( F') -S_4 ( F')= \frac{1}{30} \, F'\,(F' + 1)\,(2\,F' + 1)(2 {F'}^2 \,+  2\,F' \,+ 1)  \nonumber \\
 B &=& 4\,S_4 ( F')- 2\, F'(F'+1)\,  S_2 ( F')=\frac{1}{15 } \, (F'-1) F'\, (F' + 1)\,(2\,F' + 1) (F'+2) .
     \end{eqnarray}

We have now all we need to compute the alignment tensor    ${\cal{ A}}_{i\,j}^{e}(0 ) $, which is  expected to be 
proportional  to the traceless second-rank,  symmetric, real  tensor built from the complex vector $ \vec{b}: $
\begin{equation}
 {\cal{ A}}_{i\,j}^{e}(0 ) = - 4 \,  g_{_{F'}}^2    {\cal F}\, \left( \frac{1}{2}(b_i \, b_j^*+b_i^*\, {b }_j) -
\frac{1}{3} \vert \vec{b}\vert^2  \,{\delta}_{i\, ,j} \right),
\label{alintens}
\end{equation}
where  the angular momentum factor $ {\cal F} $  is given by:
$$ {\cal F} = A+B= \frac{1}{30}\,F'\,\left( F ' + 1 \right) \,\left( 2\,F ' - 1\right) \,
    \left(2\,F' + 1 \right) \,\left( 2\,F'  + 3 \right). 
$$
 \subsubsection{ The Stark alignment tensor in the ideal longitudinal configuration} 
We are going to apply the formula (\ref{alintens}) to the Stark amplitude which  is  always the dominant 
one in realistic  experimental situations. In the {\it ideal experiment } the applied electric field $ \vec{E}_l$
is taken colinear with the wave vector $ \vec{k} $, { \it i.e.}  $\vec{E}_l \wedge \vec{k}=0$.
  The vector $\vec{b}$ is  then  given by:
\begin{equation} 
\vec{b}_{St}= i \, \beta E_l  \hat E_l \wedge \hat \epsilon_{ex} =  i \, \beta E_l \, \hat h\, , 
\hspace {20mm} {\rm with} \hspace{10 mm} 
\hat{h}=\hat{E_l}\wedge\hat {\epsilon}_{ex}.
\end{equation}
In  this  paper  to a given vector $ \vec{v} $ there corresponds a unit vector $ \hat {v} = \vec{v}/v$ 
 with $ v= \vert \vec{v} \vert $.
 
We get then   immediately the Stark-induced alignment tensor by inserting $\vec{b}_{St}$ in Eq.(\ref{alintens}):
\begin{equation}
{\cal{ A}}_{i\,j}^{St}(0) = - 4 \,  g_{_{F'}}^2  \beta^2\,E_l^2 \  {\cal F}\, (\hat{h}_i\,\hat{h}_j -
\frac{1}{3}  \,{\delta}_{i\, ,j} ).
\label{stalin}
\end{equation}
  \subsubsection{PV effect on the Stark alignment tensor}
To get the first-order PV correction $\Delta_{pv} {\cal{ A}}_{i\,j}(0) $ to the Stark alignment tensor
we perform a first order expansion of ${\cal{ A}}_{i\,j}^{e}(0)$ by writing $ \vec{b}= \vec{b}_{St}+\Delta_{pv}\,\vec{b}
$  with $\Delta_{pv}\,\vec{b}= i {\rm Im}  \, 
E_1^{pv}\,\hat {\epsilon}_{ex} $. Let us now insert in formula
(\ref{alintens}) the  first order correction  to $ b_i\,b_j$: 
$$
\Delta_{pv}\, b_i\,b_j= \left( b_{St})_i \,\Delta_{pv} b_j^{*} + b_{St})_i ^{*}\,\Delta_{pv} b_j \right) = 
\beta \, E_l  \,{\rm Im} \,E_1^{pv}\,\left(  \hat h_i \, (\hat {\epsilon}_{ex })_j+(i
\leftrightarrow j) \right )\;. 
$$ 
We get  immediately the PV correction to the  alignment tensor:
\begin{equation}
\Delta_{pv} {\cal{ A}}_{i\,j}^e(0)=  - 4 \,  g_{_{F'}}^2  \beta^2\,E_l^2 \  {\cal F}\, {\theta}_{pv}
\left( (\hat {E_l} \wedge \hat \epsilon_{ex})_i\,(\hat{\epsilon}_{ex })_j+(i \leftrightarrow j) \right)  \;.
\label{alinpv}
\end{equation}
   As we have already noted,  the effect of the Stark-$E_1^{pv}$ amplitude is to
rotate  the alignment ellipsoid  by a small  angle $ {\theta}^{pv} $  about the direction of $ \hat{E_l } $. 
The alignment tensor  is then expected  to be subjected to the same infinitesimal rotation.   This 
result is easily verified by showing that  $\Delta_{pv}\,\vec{b}\equiv 
 i {\rm Im}\,E_1^{pv}\,\hat {\epsilon}_{ex} = {\theta}^{pv} \hat{E_l} \wedge \vec{b}_{St}$, this following immediately
from the identities:
 $$   \theta^{pv}= - {\rm Im} E_1^{pv}/ \beta E_l\;\hspace{5mm} { \rm and } \hspace{5mm} \hat{E_l}\wedge ( 
\hat{E_l}\wedge
\hat {\epsilon}_{ex})= -\hat {\epsilon}_{ex}.$$
\subsubsection{ Absence of alignment induced by the Stark-$ M_1 $ interference.}
The $ M_1$ contribution to the vector $\vec{b} $ appearing in Eq.(\ref{Teff}) is given by
$ \vec{b}_{M_1}= M_1 \hat {k}\wedge \hat{\epsilon}_{ex } $. The crucial point is that
$ \vec{b}_{M_1} $ is a purely real vector, while $ \vec{b}_{St} $ is an imaginary one. It follows 
immediately that the Stark-$ M_1 $ interference contribution to $ \frac{1}{2}(b_i \, b_j^*+b_i^*\, {b }_j)={ \rm Re} (b_i \, b_j^*) $
 vanishes since $ (b_{St})_i \, ({b}_{M_1})_j^*+({b}_{M_1})_i \,(b_{St})_j^* $ is clearly imaginary. In conclusion,
{\it the Stark-$ M_1 $ interference term does not contribute to the alignment tensor 
within the hypotheses leading to Eq.(\ref{alintens}) }. A crucial one of these concerns the $6S,F$ density matrix which is
assumed to be a thermal distribution: $\rho_g  \propto  P_F$. This property  does not hold in presence of a static magnetic
field 
$ \vec{B} $, if the excitation beam intensity varies significantly within a frequency interval of the order of the Zeeman
splittings. We will show later on, that if $ \vec{E} $ has a component transverse to $\vec{k}$,  
then  the Stark-$ M_1 $ interference leads to a non-zero alignment contribution.

It is worth stressing the importance of the result obtained in this section. The $M_1$-Stark interference is a potential source of
systematics which has required special care in all experiments performed on highly forbidden transitions in the transverse
field configuration so far
\cite{bou82,woo99}.  Thus the absence of
$M_1$-Stark interference in the longitudinal field configuration examined here is a significant advantage.
In this configuration, it is clear that the $\pi/2$ phase difference between the $M_1$ and the Stark amplitudes prevents 
there being any 
interference whatever the excitation polarization,  and this even if the atomic system is perturbed by a magnetic field.
 
  It is only in presence of a stray
{\it transverse} electric field that we shall have to consider the possibility of systematic effects associated with the magnetic
dipole amplitude. 
 
\subsection{Optical anisotropy resulting from the APV alignment}
While the population inversion causes an amplification of the probe beam,  its
polarization is altered by the angular anisotropy created in the excited state. We obtain the {\it atomic anisotropies} in the
excited state by detecting the {\it optical anisotropies that they induce on the amplified probe beam}. Here the vapor is
endowed with an atomic alignment. Associated with the alignment ellipsoid is an ellipsoid of refractive index for
light resonant with one hyperfine component of the $7S-6P_{3/2}$ transition. The relative magnitude of the axes
depends on the hyperfine component. The eigenaxes are those of the ellipse resulting from the  intersection of the alignment
ellipsoid by a plane orthogonal to the wave vector
$\hat k $ of the probe beam. The imaginary part of the refractive index is responsible for the gain which takes two different
values depending on whether the linear polarization $\hat
\epsilon_{pr}$ is aligned along one or the other eigenaxis, an effect generally dubbed {\it linear dichroism}, while the
real part is responsible for {\it birefringence} of the vapor. 
For the probe beam which propagates along $\hat z$, the Stark and the PV alignments both induce a linear dichroism, but
the former with axes along $\hat \epsilon_{ex}$ and $\hat E_l \wedge \hat \epsilon_{ex}$ and the latter with axes $\hat X$ and
$\hat Y$. It is this latter contribution, the optical anisotropy resulting from the APV alignment, which gives rise to the APV signal in
the pump-probe  experiment. As a consequence the global gain matrix of the excited vapor has its axes tilted with respect to
$\hat
\epsilon_{ex}$ by the tiny angle $\theta^{pv}$. The tilt is opposite for two mirror-image configurations associated with
opposite signs of the pseudoscalar  
${\cal S}^{ch} = (\hat \epsilon_{ex} \cdot \hat
\epsilon_{pr})(\hat \epsilon_{ex} \wedge \hat \epsilon_{pr} \cdot \vec E_l)$. 
 Hence the notion of a {\it chiral optical gain}.
The probe polarization aligned along
$\hat \epsilon_{ex}$ at the cell entrance actually does not lie along an eigenaxis. Consequently its direction is modified
during the propagation of the probe beam through the vapor. This modification is measured using polarimetry methods
which allow us to obtain an absolute determination of $\theta^{pv}$, as explained in the next section.
\subsection{Absolute calibration of the tilt angle $\theta^{pv}$}
The physical interpretation of APV given in  \S~2.1 suggests a natural method of calibration. The APV
contribution can  be seen as resulting from the dominant PC contribution via an infinitesimal rotation of angle $\theta^{pv} $
about $\hat E_l$. {\it By performing small rotations of the excitation polarization} by a known angle
$\theta_{cal}$, while keeping
$\hat
\epsilon_{pr}$ unchanged at the input, we can induce an  optical
anisotropy for the probe exactly similar to the APV one, eventhough it is P-conserving. The ratio of
these two effects, measured  under identical conditions and distinguished by their opposite dependence on $\theta_{cal}$
and $E_l$ reversals, yields directly
$\theta^{pv}/\theta_{cal}$. By measuring the ratio, it is possible to eliminate different effects affecting the
magnitude of each signal individually which would otherwise be difficult to predict quantitatively (see \S~3.2).
\section{ Balanced-mode polarimetry measurements on the amplified probe beam}
\subsection{Operating conditions providing polarimeter imbalances insensitive to the PC alignment}
Since we want to measure a small PV anisotropy in presence of a large PC one, we orient 
the polarimeter so that it is insensitive to the latter, and fully sensitive to the former. We have
chosen a polarization beam splitter cube having its axes $\hat X$ and $\hat Y$ oriented at $\pm 45^{\circ}$ to the
excitation and probe polarizations. The input probe polarization is a superposition of the two eigenpolarizations,
$\hat\epsilon_X
$ and $ \hat \epsilon_Y$ of the polarimeter: 
\begin{equation}
 \hat \epsilon_{pr} = (\hat\epsilon_{X} + \hat \epsilon_{Y} )/\sqrt{2} .
\end{equation}
In absence of the excitation
beam, the electronic gains of the two channels are adjusted to ensure equality of the probe signals detected in both
channels,
$S_X  = S_Y$. However, due to possible drifts in the electronic or optical components, one has to isolate the true atomic
contribution to the imbalance. For each excitation laser pulse, the imbalance for the amplified probe pulse,
$D^{amp}=
\frac{(S_X-S_Y)}{(S_X+S_Y)}$ is compared with a reference value $D^{ref}$, measured when all the $7S$ excited atoms have
decayed. Thus {\it the doubly differential signal},
$D^{amp} - D^{ref}$, selects at the laser repetition rate a true atomic effect, $D^{at}$, free of the polarization defects
present on the probe beam path. 
\subsection{Relation between atomic alignment and polarimeter imbalance }
Since the probe polarization is adjusted at cell entrance to be parallel to $\hat \epsilon_{ex} = \hat y$ which (without APV)
defines the eigenaxes of the vapor gain, it should remain parallel to one eigen axis of the gain during
propagation through the vapor.  An atomic imbalance is expected to arise specifically from a difference of light amplification for a
probe beam polarized along $\hat X$ or $\hat Y$.

In a previous publication (see \cite{lin89}  Eq. 20 or \S~5.2) 
we have calculated the intensity and polarization modification of a linearly
polarized probe beam resonant for a single hyperfine component of the probe transition $7S_{1/2} \rightarrow 6P_{3/2}$.
 Following a similar approach, we shall write the  light intensity  difference   $S_X - S_Y $  as an expression 
proportional to the contraction of the  alignment tensor  $ {\cal{ A} }_{i\,j}^e(\tau) $ - at the time of detection - with the traceless
symmetric tensor
$  {\cal{ D} }_{i\,j}$  which  characterizes the differential  two-channel polarimeter:
\begin{equation}
 {\cal{ D} }_{i\,j}= ({\hat\epsilon}_{X} )_i\,({\hat\epsilon}_{X} )_j
-({\hat\epsilon}_{Y} )_i\,({\hat\epsilon}_{Y} )_j. 
\label{dettensor}
\end{equation}  
Let us denote by $ \vec{D} $ the electric dipole moment operator and by ${\vec{d} }^{~se}$  the effective dipole operator  
which induces specifically the stimulated emission transition  $ 7 S,F'  \rightarrow  6P_{3/2}, F_p $:
$$ {\vec{d} }^{~se} =     P_{_{F_p}}  \,\vec{D} \,  P_{_{F'}}$$ where $P_{F_p}$ 
is the  projection  operator  on the $6P_{3/2}, F_p$ sublevel while $P_{_{F'}}$ is relative to the $7S,F'$ sublevel.  (Note that  $ 
{\vec{d} }^{se}$  is non-hermitian,  but still  a  vector operator.) Now let us  introduce the second-rank tensor
operator:
 $ {\cal{O}}_{i\,j}= ({d} ^{se}_i )^{\dagger}\, {d}^{se}_j $ whose action is restricted to the  $ 7 S, F' $ subspace.
Applying the Wigner-Eckart theorem, we can write:
\begin{equation}
{\cal{O}}_{i\,j}= {(d} ^{se}_i)^{\dagger} \, {d}^{se}_j= C_{se}  (6P_{3/2}\,  F_p \vert  7S \, F' ) \, F_i \,F_j \; P_{_{F
\prime}} \;,
\label{seop}
\end{equation}
where $C_{se}  (6P_{3/2}\,  F_p \vert  7S \, F' )$ is a real constant (explicitely calculated in \cite{lin89}).

 We now have all the tools to obtain the result announced above.
\begin {eqnarray}
S_X - S_Y &\propto&  {\rm  Tr}
\{ \rho_e (\tau)\,( {\vec{d} ^{se}}\cdot \hat\epsilon_{X})^{\dagger} \, ( {\vec{d} ^{se}}\cdot \hat\epsilon_{X})  
-( X\rightarrow   Y ) \} \nonumber \\
 &\propto& {\cal{ D} }_{i\,j}\cdot {\rm Tr}\{ \rho_e(\tau)\, {\cal{O}}_{i\,j}\}  \propto
 {\cal{ D} }_{i\,j}\cdot {\cal{A}}_{ij}^e(\tau) \,,
\label{imbali}
\end{eqnarray} 
 where the dot indicates that the two tensors have to be contracted, {\it i.e.} summations have to be performed 
upon the repeated indices. 

Let us calculate the imbalance signal  $S_X - S_Y $ using the  alignment tensors $  {\cal{A}}^e_{i\,j}$ 
 derived  in the previous section and given in equations (\ref{stalin}) and (\ref{alinpv}).
 We deal first with the  Stark-induced alignment:
$$  
S_X - S_Y \propto  {\cal{ D} }_{i\,j}\cdot  {\cal{ A}}_{i\,j}^{St}(0 ) =
 - 4 g_{_{F'}}^2 \,{\beta }^2  E_l^2  {\cal F}\, {\cal{ D} }_{i\,j}\cdot   {\hat{h} }_i \, {\hat{h} }_j=  
- 4 g_{_{F'}}^2 \,{\beta }^2  E_l^2  {\cal F}\left ( (\hat\epsilon_{X}\cdot \hat{h}) ^2-(\hat\epsilon_{Y}\cdot \hat{h}) ^2
\right) =0 \,.
$$
 The above cancellation follows immediately from the two-channel
polarimeter  eigenaxis  expressions  in a defect-free experimental set up :
 \begin{equation}
\hat\epsilon_{X,Y}= \frac{1}{\sqrt{2}}(\hat \epsilon_{ex} \pm  \hat \epsilon_{ex}^{^{\,\perp}}) \; \hspace{15mm} \;   {\rm 
with }\hspace{8mm}
 \hat \epsilon_{ex}^{^{\,\perp}} = \hat z \wedge \hat \epsilon_{ex} \,,
\end{equation}
$\hat z $ being a direction colinear to $\hat k$ and $\hat E_l$ which remains fixed in the laboratory whatever parameter
reversal is made. This obviously implies: $ \hat\epsilon_{X}\cdot\hat{h}= -\hat \epsilon_{Y}\cdot \hat{h} $.

In absence of defects, such as deliberate or accidental optical misalignements, stray electric and magnetic fields - to be
discussed extensively in the next section - the  only possible source for non-zero  polarimeter imbalance  will turn out to be
the PV alignement calculated in the previous section (see Eq.(\ref{alinpv} )).  Let us insert   $ \Delta_{pv} {\cal{
A}}^e_{i\,j}(0) 
$ in the formula  (\ref{imbali})  giving $ S_X - S_Y$  in terms of the alignment tensor:
\begin{eqnarray}
(S_X - S_Y)_{pv}  &\propto &  {\cal{ D} }_{i\,j}\cdot \Delta_{pv} {\cal{ A}}^e_{i\,j}(0) =
  - 4 g_{_{F'}}^2 \,{\beta }^2  E_l^2  {\theta}_{pv} {\cal F }\,{\cal{ D} }_{i\,j} \cdot \left( \hat{h} _i (\,{\hat \epsilon¼}_{ex} )_j
+ (i
 \leftrightarrow j) \right)   \nonumber \\
&=&  - 4 g_{_{F'}}^2 \,{\beta }^2  E_l^2  {\theta}_{pv}\, {\cal F }\left ( (\hat{h}\cdot \hat\epsilon_{X}) \,(\hat
\epsilon_{ex}\cdot\hat\epsilon_{X}) - (X\rightarrow  Y  ) \right)   \nonumber \\
&=& - 4 g_{_{F'}}^2 \,{\beta }^2  E_l^2  {\theta}_{pv}\, {\cal F }\left ( (\hat{E_l}\wedge \hat\epsilon_{ex} \cdot
\hat\epsilon_{X})
\,(\hat
\epsilon_{ex}\cdot  \hat\epsilon_{X}) - (X\rightarrow  Y) \right) \,.
\end{eqnarray} 
In the above formula  appears, as expected,  pseudo-scalars built from the physical vector objects involved in the
experiment namely the longitudinal field $\vec E_l$, the laser polarization $\hat \epsilon_{ex}$ and
the polarimeter eigenaxes:  
\begin{equation}
  {\cal S }_{chir} (\hat\epsilon_X, \hat E_l)= (\hat{E_l}\wedge\hat\epsilon_{ex}\cdot \hat\epsilon_{X}) \,(\hat
\epsilon_{ex}\cdot\hat\epsilon_{X})  = -{\cal S }_{chir} (\hat \epsilon_{Y}, \hat E_l) \;.
\label{chir}
\end{equation}
 Note that the last equality reflects the fact that  we obtain  
$ \hat\epsilon_{Y}$ from $\hat\epsilon_{X} $ by performing a mirror reflexion with respect  to the plane
 containing $ \hat\epsilon_{ex} $ and $\vec{E_l} $ .

All of this means that the imbalance $D^{at}= (S_X - S_Y)/(S_X + S_Y)$  allows one to search for the tiny PV effect using
dark field detection. {\it In a general way such an atom-induced imbalance should vanish unless some chirality is
present in the experiment}.

In Eq.(\ref{imbali}), $\tau$ is the instant of detection which differs from the instant of excitation taken for origin of time. This
means that in presence of a magnetic field we have to take into account the evolution of the excited state over the time interval
 which separates the instants of excitation and stimulated emission.  

We can write the general relation:
\begin {equation}
 D^{at}= K \; {\cal{ D} }_{i\,j}\cdot {\cal{A}}_{ij}(\tau) \,. 
\label{atimb}
\end{equation}
Several effects are embodied in the evaluation of the proportionality coefficient $K$: not only transition
probabilities  depending on the hyperfine quantum numbers of the hyperfine states
connected by the probe transition \cite{lin89}, but also the probe pulse time and duration,  the optical thickness of the vapor
at the probe laser wavelength as well as saturation effects  induced  by the probe beam
\cite{bou96}. The calculation of $D^{at}$ involves the problem of the amplification of the probe while it propagates
through the vapor. This has been considered previously both theoretically \cite{bou96} and experimentally
\cite{cha98}. We refer the reader to previous work 
 for the evaluation of the amplification factor common to both  $S_X$ and $S_Y$ as well as the
asymmetry amplification which affects directly the difference $S_X - S_Y$ and thereby the atomic
imbalance
$D^{at}$.  {\it The crucial point here is that our measurement method includes a calibration procedure allowing us to 
eliminate K by performing ratios between atomic imbalances of different physical origin but similar optical
properties}. To do this we apply a small rotation of
  $\hat \epsilon_{ex}$ of angle $\theta_{cal}$ around $\hat z$ before entrance in the cell, {\it the sense of this
rotation being independent of the direction of $\hat E_l$} . The calculation of the corresponding  correction  to the
alignment tensor  mathematically is  identical to that of $
\Delta_{pv} {\cal{ A}}_{i\,j}(0) $  except for  the trivial replacement $ \theta^{pv} \hat E_l \rightarrow \theta_{cal} \, \,
\hat
 z $. This yields the atomic imbalance:
\begin{equation}
 D^{at}(\theta_{cal} {\rm -}odd\,, E_l {\rm -}even) = - 8 \,  g_{_{F'}}^2  \beta^2\,E_l^2 \, \theta_{cal} \,  K \; {\cal F}\;   
{\cal S }_{chir} (\hat\epsilon_{X}, \hat E_l) \, ( \hat z \cdot \hat E_l ) \,, 
\label{calimb}  
 \end{equation}
with which we compare the atomic APV imbalance:  
$$D^{at}(\theta_{cal}{\rm -}even\,, E_l {\rm -}odd)  = - 8 \, 
g_{_{F'}}^2  \beta^2\,E_l^2 \, \theta^{pv} \,  K \;  {\cal F}\;   {\cal S }_{chir} (\hat\epsilon_{X}, \hat E_l) , $$  allowing us to
calibrate $\theta^{pv}$ in terms of the known angle $\theta_{cal}$. All the atomic factors embodied in $K$,  as well as
$E_l^2$ are eliminated in the ratio of both imbalances, {\it provided they are measured under identical conditions}.
As long as this important condition is fulfilled, the relation
\begin{equation}
\frac {D^{at}( \theta_{cal}{\rm -}even\,, E_l {\rm -}odd) }{D^{at}(\theta_{cal} {\rm -}odd\,, E_l {\rm -}even)} =
\frac{\theta^{pv}}{\theta_{cal}} \, ( \hat z\cdot \hat E_l )
\label{Datpvcal} 
\end{equation} 
{\it holds whatever the detection conditions, for any hyperfine component}. 
 The factor $(\hat z \cdot \hat E_l) =\pm 1$, depending on the sign of the projection of $\vec E_l$ on the
fixed direction $\hat z$, will appear frequently in the forthcoming formulae.  
\subsection{Symmetry breaking associated with a $B_z(E_l-odd)$ magnetic field}
 Since it is a pseudo-vector, a longitudinal magnetic field,breaks the mirror reflexion symmetry with respect to the plane
containing
$\hat \epsilon_{ex}$ and $\vec E_l$. It does not however break the cylindrical symmetry of the experiment. Therefore, it
escapes the methods of diagnosis based on global rotations of the experiment, discussed in the present paper. This effect
discussed previously
\cite{gue02} is quoted here just for completeness.

A magnetic field $B_z$ causes Larmor precession of the angular momentum
$\vec F$ about the $\hat z$ axis for the average duration
$\tau$ which separates the time of stimulated emission from the time of excitation. Consequently, 
after evolution of the density matrix in the excited state, the axes of the Stark alignment ${\cal A}_{i
\, j}(\tau)$ no longer lie in the symmetry planes. They deviate by an angle $\gamma B_z  \tau$ (with $\gamma B_z =
\omega_{_{F'}}$ being the angular precession frequency). This effect is exactly similar to the calibration effect provided the
angle $\theta_{cal}$ be replaced by $ \gamma B_z  \tau$.  A problem arises only if the
$B_z$ field is odd under $\vec E_l $ reversal, in which case there arises a signal simulating exactly the PV effect. 
To account for it, we measure the magnitude of the $B_z$-odd field seen by the atoms.   This
 is performed by measuring the optical rotation (see \S~4.3.4) induced by $B_z$-odd on a probe transition which is
particularly sensitive $ 7S, F=4 \rightarrow 6P_{3/2}, F=5$ \cite{bou95}.  In our past  measurements \cite{bou02} it 
has vanished part of the time and never exceeded  50
$\mu$G for the rest of the time. Then, we can correct for its effect on the PV signal.

\subsection{Optical rotation detection}
In the polarimeter configuration chosen here, the imbalance is not only sensitive to linear dichroism relative to the $\hat X$,
$\hat Y$ axes but also to an optical rotation \cite{gue97}. Consequently, a contribution to the atomic imbalance
may arise from  an atomic orientation in the 7S state, ${\cal P}(\tau) = {\rm Tr}\{\rho_e(\tau) \,\vec F \cdot \hat z \}$.
Strictly speaking it corresponds to an additional term in Eq.(\ref{atimb}) (see \cite{lin89}) which we have omitted here for
two reasons: i) it vanishes in the ideal APV configuration; in real conditions, only combination of
imperfections can lead to a non-zero signal; ii) instead of a linear dichroism effect, this corresponds to an optical rotation of
the linear probe polarization, {\it i.e.} an optical effect which our method of measurement allows us to discriminate
unambiguously against linear dichroism. The latter causes the probe polarization to rotate towards the axis of larger gain,
while the sense of the  rotation imposed by an optical activity is the same whatever the direction of the input polarization.
Therefore, in order to distinguish between both effects, we use their opposite behavior when the direction of the probe
polarization is switched  in the vapor cell, with a half wave plate $(\lambda /2)^{pr}$, from
$\parallel$ to $\perp \hat \epsilon_{ex}$, configurations respectively dubbed para and ortho. By rotating the 
probe polarization through 90$^{\circ}$ at cell entrance and by leaving the polarimeter and the excitation
polarization unaltered, we switch the probe polarization from one bisector of the polarimeter eigenaxes to the other.
Hence, the polarimeter still operates in balanced mode and the imbalance associated with the  linear dichroism effect, once
calibrated, remains identical in magnitude and sign. The effect of optical rotation on the other hand changes sign 
and can be rejected \cite{gue97}.

With this method, the probe polarization is rotated by 90$^{\circ}$ with respect to the polarimeter axes. Another possibility
consists in applying a second 90$^{\circ}$ rotation to the probe polarization {\it outgoing} from the cell. Without tilts induced
by the vapor, the probe polarization analyzed by the polarimeter in the ortho configuration recovers the same direction as that
it has in the para one. This method {\it a priori} equivalent to the former one, has the advantage to make easy the use of a
polarization magnifier, a dichroic component \cite{cha97} which would otherwise come into conflict with measurements
involving probe polarization rotations. 
 \subsection{Global rotation of the experiment about the beam axis}
As explained previously, the balanced-mode operation of the polarimeter, favoured for
optimal detection of small asymmetries,  requires that at the cell entrance  we choose either a para or 
an ortho configuration. When we rotate $\hat \epsilon_{ex}$ by
$45^{\circ}$ increments, we simultaneously rotate $\hat \epsilon_{pr}$ by the same angle. We do this for both the para
and the ortho configurations.  Thus, eight different pump-probe configurations (four para and four
ortho) are used for the measurements.  In principle, when we rotate $\hat \epsilon_{ex}$ we should rotate the polarimeter
in order to ensure that its eigenaxes remain at $\pm 45^{\circ}$ angle from $\hat
\epsilon_{ex}$, the axes of the main Stark alignment to which we want to remain insensitive. But this operation 
is hardly compatible with excellent mechanical stability. We have found that it can be advantageously replaced by a second 
45$^{\circ}$ rotation of $\hat \epsilon_{pr}$ at the cell exit, opposite to that performed at the entrance. All
polarization rotations are performed by insertion and removal of half-wave plates placed on each input beam and before
the polarimeter on the probe emerging from the cell. The {\it output} plates 
are oriented and controlled in a way which allows the analyzed probe polarization, in absence of tilts induced 
by the vapor, always to lie along the same bisector of the polarimeter eigen-axes, {\it whatever the input
configuration of the $\hat \epsilon_{ex}, \hat \epsilon_{pr} $ polarizations}. 

We can verify that this operation, where the cell is sandwiched between two half wave plates having their axes parallel, is
physically equivalent to rotating the input polarization and the polarimeter by the same angle as that carried out by one half
wave plate. Let us introduce the transformation T produced by the vapor cell whose effect on $\hat
\epsilon_{pr}$we want to measure and the transformation applied by insertion of a half wave plate ${\cal S}_{\delta}$,
which is a symmetry with respect to the plate axes  of direction $\delta$. With the plates absent, the polarizations at the
input and output of the cell are: $\hat \epsilon_{out} = T \hat \epsilon_{in}$; with the input and output plates inserted,
$\hat
\epsilon'_{in} = {\cal S}_{\delta} \hat
\epsilon_{in}$ at cell entrance, while at entrance of the polarimeter $\hat \epsilon'_{out} = ({\cal S}_{\delta} T {\cal
S}_{\delta} ) \hat \epsilon_{in}$ .  We note that the scalar product is conserved in the operation ${\cal S}_{\delta}$ and that   
$({\cal S}_{\delta})^{2} =  1 \hspace{-0.8mm}{\rm I}$.  The measured quantity in each channel is the square of the scalar
product $\hat X \cdot \hat \epsilon'_{out}
\equiv \hat X \cdot ({\cal S}_{\delta} T {\cal S}_{\delta} ) \hat \epsilon_{in} \equiv ({\cal S}_{\delta} \hat  X) \cdot (T
\; {\cal S}_{\delta} \hat \epsilon_{in}).$ Hence, we measure the same quantity as if we had applied the same
transformation to the polarimeter and to the input polarization.

 {\it To simplify the following discussion we suppose hereafter that the polarimeter is rotated in the same way as  the probe
polarization and that the axes $\hat x, \hat y, \hat X$ and $\hat Y$ are  linked to the input polarizations and to 
the polarimeter in their global rotation about the beam axis.}
During the rotations, the two components of the probe polarization along the eigen-axes of the polarimeter,
$\hat\epsilon_X, \hat \epsilon_Y$, remain at $\pm 45^{\circ} $ to the input probe polarization, both in para and ortho
configurations.  

To be complete, let us mention that we insert one more half-wave plate, $(\lambda /2)^{det}$, just in front of the
polarimeter which performs a symmetry of the outgoing probe polarization with respect to the symmetry plane of the
polarimeter, for discrimination between {\it true polarization tilts} and instrumental imbalances \cite{gue97}. Then, using
the four reversals $\sigma_{cal}= \pm 1 ,\, \sigma_{E_l}= \pm 1 ,\, \sigma_{pr}= \pm 1 ,\, \sigma_{det}= \pm 1 , $
of the calibration angle $\theta_{cal}$, the electric field $\vec E_l$ and half-wave plates  $(\lambda /2)^{pr}$ and 
$(\lambda /2)^{det}$ respectively, the PV quantity $\theta^{pv}$ is reconstructed as the $E_l $-odd, calibrated linear
dichroism: 
\begin{equation}
G = \theta_{cal}  \left <    \hspace{-1mm} \sigma_{E_l}    \hspace{-1.5mm}  \left  [\frac{ < \sigma_{det} D^{at}
(\lbrace \sigma_j  \rbrace) >_{\sigma_{det} \sigma_{cal}}}{<  \hspace{-1mm} \sigma_{det} \sigma_{cal}D^{at}(\lbrace \sigma_j  
\rbrace )
\hspace{-1mm}> _{\sigma_{det}
\sigma_{cal}}}
\right ]   \right >_{\hspace{-1mm}\sigma_{E_l} \sigma_{pr}} \,.
\label{dichro}    
\end{equation}
The experiment  \cite{bou02,gue98} consists in recording successively the four values of $G_y, G_x, G_Y, G_X$ of G for the
four configurations $\hat y, \hat x, \hat Y, \hat X$ of $\hat \epsilon_{ex}$.
In the next sections of this paper, we show that the cylindrical symmetry can help in identifying and rejecting contributions
to $G$ (Eq.(\ref{dichro}))
due to transverse $\vec E$ and $\vec B$ fields, or misalignments.
\subsection{Defect invariance under global rotations of the experiment}
We cannot take for granted that during global rotations of the experiment about the common beam axis  all transverse fields
remain invariant. In this section we want to discuss this assumption and present arguments which give it some support.
There is no problem of course concerning  the laboratory magnetic field imperfectly compensated, nor a possible
transverse
electric field which results from an $\hat E_l / \hat k$ misalignment and reverses with the  voltage applied to the cell. But one
may be concerned by the behavior of the electric and magnetic fields arising from the space charge developped inside the cesium
cell. The space charge results from  photoionization taking place on the cell windows at each excitation pulse \cite{gue02}.
Therefore, the question at issue here  is that of a possible correlation between the direction of the excitation polarization and the
electron distribution responsible for transverse $\vec E_t$ and $\vec B_t$ fields. Would such a correlation
exist, our assumption would fail.  
Thus, we are concerned by the  angular    
 distribution of the electrons  emitted by the surface of a sapphire crystal 
normal to the trigonal symmetry axis via the photoionization  process. The 
linearly polarized photons, with $ \lambda= 540 $ nm,  propagate along the symmetry axis.
In practice, the sapphire surface, at a temperature around 500 K, constitutes the internal side of  the windows of the 
 cylindrical cell containing cesium vapor. (This latter is at saturated vapor pressure of a reservoir kept at
410 K). Those windows have been annealed at a temperature of $\sim 1400$~K before being mounted on the
cylindrical tube. After annealing the surface is reconstructed: we have found by AFM imaging and laser beam diffraction 
\cite{lin02} the
presence of regularly spaced vicinal steps, several hundreds of nanometers apart.  This situation is known to be favorable for
reducing the adsorption probability, except for edge sites which favor chemisorption \cite{zav95} but concern only a tiny fraction
of the surface.    

  According to the findings   of references \cite{cirac83} and \cite{rod96}, we shall assume 
that the surface states  can be constructed mainly  from the Al  $ 3p $   orbitals. 
We shall take for the density matrix describing the  surface states the following expression:
\begin{equation}
{\rho}_S=\sum_i  a_{t} (i) \, \left ( \vert 3 p_{x}(i) \,\rangle \langle 3 p_{x}(i)\vert +
\vert 3 p_{y}(i) \,\rangle \,\langle 3 p_{y}(i)\,\vert  \right)+
 a_{l} (i)   \vert 3 p_{z}(i) \rangle \langle  3p_{z}(i)\,\vert  
\label{roS}
\end{equation} 
The summation   $i$ runs over the surface aluminum  atoms endowed with dangling $ 3p $  bonds due 
to the presence of cesium adatoms. Under  the experimental working conditions, T=500K, the 
surface coverage is small. 
It is then reasonable to assume that the dangling  Al  $ 3p $  bonds are 
randomly distributed  on the surface so that the phase factor between the orbitals  $(i)$ and $(j)$,
 $\, \exp(i \,(\vec{r}_i-\vec{r}_j)\cdot\vec{k}) $  is also randomly distributed. This explains why
we have not included in $ {\rho}_S $ non diagonal terms like $\vert 3 p_{x}(i) \,\rangle \langle 3
p_{x}(j)\vert $. 

Since the incident photons belong to the optical  range it is legitimate to use the dipole approximation for describing the
photoionization process.  Centrifugal 
 barrier effects  imply that  the $ p\rightarrow s$ transition amplitude dominates the  $ p\rightarrow  d $  one. Let us denote 
by $  x\,{\psi}_{3p}(r), y \,{\psi}_{3p}(r) $ and $ {\psi}^{out}_s (p_e,r)  $ the  wave functions associated respectively  with  
the $3 p_{x}$,  $3 p_{y}$ orbitals and the outgoing  $s$ wave  electrons  with momentum $p_e$. With the help
 of  equation  (\ref{roS}) the surface  photoionization cross section   is given, up to phase space  factors, by:
\begin{equation}
  \sigma( 3p \rightarrow s ,\, p_e ) \propto \vert\int \, d^3r  \;  x\,{\psi}_{3p}(r) ( \vec{r} \cdot \vec\epsilon_{ex} )\, 
 {\psi}^{out}_s (p_e,r) \vert^2 + \left( x \rightarrow  y      \right)
\label{sigion}
 \end{equation}
 Remembering that the photon momentum is along the $z$ axis, we can write $ \vec{r} \cdot  \hat \epsilon_{ex}= 
x  \cos(\phi) +y \sin(\phi), $  where $ \phi  $  gives the direction  of  the polarization vector $  \vec \epsilon_{ex} $ in the $
x,y$ plane.  
 Performing the space integral  in equation (\ref{sigion}) one sees that  the first term in the right hand side is 
proportional  to $\cos(
\phi)^2  $ and the second  to $ \sin(\phi)^2  $ times the square of the same radial integral $ \int_0^\infty dr\,r^4
\,{\psi}_{3p}(r)\, {\psi}^{out}_s (p_e,r) $. In more physical   terms , it means that the distribution  of the photo-electrons has
no  dependence on the
direction of the photon  linear polarization  $\vec\epsilon_{ex} $ in the $ x,y$ plane.  Such a result hinges on the
assumption  that  the surface states near the edges of the conduction band  are produced mainly by  Al  $3p_{x,y}$  orbitals,
according 
 to the authors of reference \cite{cirac83}, section IV.A. It should be kept in mind that adding a contribution
 $  \propto \vert 3 s(i)\,\rangle \langle 3s(i)\vert $ to the surface density matrix  ${\rho}_S $ would indroduce a $ \cos(\phi
)^2$ dependence in the angular distribution of the photo-electrons \footnote{This can be verified by making in Eq.
(\ref{sigion}) the substitutions needed for the evaluation of $\sigma(3s \rightarrow p,\,p_e)$ instead of 
$\sigma(3p \rightarrow s,\,p_e)$, namely $ x\,{\psi}_{3p}(r)
\rightarrow
\psi_{3s}(r)$ and  $ {\psi}^{out}_s (p_e,r)
\rightarrow (\vec p_e \cdot \vec r) \; {\psi}^{out}_p ( p_e, r)$  leading to an angular distribution of $\sigma(3s
\rightarrow p,\,p_e)
\propto \vert \hat \epsilon_{ex} \cdot \vec p_e \vert ^2$, hence to a $\cos(\phi)^2 $ dependence in the transverse plane.   
}, like for instance  in the photelectric effect in atomic hydrogen. 
No evidence for a  $ \cos(\phi )^2$  term  has been found  in the  measured radial electric field produced by the  beam  of 
surface photo-electrons accelerated by an applied electric static field normal to the surface \cite{gue02}. This gives some
empirical support for  the basic assumptions leading to formula (\ref{roS}), in particular, for the absence of Al $3s$ orbital
contributions.

\section{ Cylindrical  symmetry breaking  generated 
 by transverse $\vec E_t$, $\vec B_t$ fields or misalignment.}
In this section, we consider the modification of the the two-channel polarimeter imbalance $D^{at} $ resulting from the
simultaneous presence of stray  transverse $\vec E_t$, $\vec B_t$ fields. The important  feature of the present mechanism is
that it    can generate a PV-like signal which, although a physical scalar,  shares the properties of the true PV signal 
 and in particular the cylindrical  symmetry. Fortunately, {\it this  potential source of  systematic effect  is accompanied 
by  another imbalance contribution which breaks the cylindrical  symmetry}. In the present deviation  
from the {\it ideal configuration}, the imbalance signal  anisotropy indicates the presence of  possible systematic  effect 
and  can yield an upper bound for its magnitude. 
 \subsection{Alignment  correction induced by   $\vec E_t$ and $\vec B_t$ fields. }
 Now, we evaluate the second-order correction to the alignment tensor $\Delta^{^{(2)}} {\cal A}_{i\,j}^e $ 
generated by the combined action of the fields $\vec E_t$ and $\vec B_t$, using the method 
developed in section 2.2.1.
\subsubsection{$\vec E_t$-induced correction to the Stark alignment.}
We start by the first-order modification of the Stark alignment    $\Delta^{^{(1)}} {\cal A}_{i\,j}^e(0) $   
resulting from the presence of  $\vec E_t$. The vector $ \vec{b}_{St} $   which defines the Stark  transition matrix 
has a correction term $\Delta_{E_t}^{^{(1)}}\vec{b} = i\, \beta \, \hat{E_t}\wedge \hat \epsilon_{ex}$, 
which clearly lies along $\hat{k}$. The corresponding correction $\Delta^{^{(1)}} {\cal A}_{i\,j}^e(0) $   is  then proportional
to $( \vec{b}_{St})_i\, \hat{k}_j +(i  \leftrightarrow j)$. It does not contribute to the polarimeter imbalance $ S_X -S_Y \propto 
{{\cal D}_{i\,j} \cdot \cal A}_{i\,j}^e $ since in the {\it ideal configuration } the detector tensor is 
purely transverse,  {\it i.e.} $ \hat{k}_i \cdot {\cal D}_{i\,j} =0.$
\subsubsection{ $\vec B_t$ Larmor   precession  of the $\vec E_t$ correction to the Stark alignment}
 For there to be a non-zero correction, another kind of defect must be present. For  instance a 
 transverse magnetic field $\vec B_t$ would induce a  Larmor precession of the tensor $ {\cal A}_{i\,j}^e(0)  $ 
(Eq.(\ref{alintens})) making the defect detectable by the polarimeter. The Larmor precession, of angular frequency
$\omega_{_{F'}}$ is equivalent to a  rotation of the vector  
$\vec{b}$ (and $\vec{b}^*$) about the axis $ \hat{B}_t$  by the   angle 
 $\omega_{_{F'}} \,\tau= \gamma B_t \tau $. The duration   
  $\tau$  represents the averaged time spent by  the atoms in the excited state with typical values for
short pump-probe delay between 5 and 15 ns,
depending mainly on the saturation. Note that the sign of the Larmor precession about
$\vec B_t$, given here by the sign of $\omega_{_{F'}}$ is opposite in the two $7S$ hyperfine states \footnote{ Given that
the magnetic moment of the electron is negative, the precession occurs in the positive sense for the $F' = I + 1/2 $ sublevel
with  $g_{_{F'}}>0$.}. In practice $\vert \omega_{_{F'}}\vert \,\tau
\ll 1$,  so that the rotation can be considered as infinitesimal, and the variation of $ \vec{b} $   induced by   the Larmor
precession  can be written as: $ \Delta_L ^{^{(1)}}\vec{b} = \gamma B_t \tau  \hat{B}_t \wedge \vec{b} $. The combined
effect of 
$\vec E_t$ and $\vec B_t$ is then described by a second-order correction to the vector $\vec{b}_{St}$:
\begin{equation}
  \Delta^{^{(2)}} \vec{b_{St}}= \Delta^{^{(1)}}_L ( \Delta^{^{(1)}}_{E_t} \,\vec{b_{St}} )= 
i \beta \, \gamma \,\tau \, \vec{B}_t \wedge (\vec{E}_t \wedge \hat \epsilon_{ex} )=
i \beta \, \gamma \,\tau \left ((\vec B_t \cdot \hat\epsilon_{ex})\vec{E_t} - (\vec{B}_t \cdot \vec{E_t}) \hat \epsilon_{ex}
 \right) . 
\end{equation}
We obtain the   mixed    second order  $\vec E_t\, ,\vec B_t $ correction to $ {\cal{ A}}_{i\,j}^{e}(\tau)$ contributing
to $S_X - S_Y$ by
replacing in Eq.(\ref{alintens}) $ {\rm Re} \{b_i \, b_j^* \} $ by $  {\rm Re} \{ (\vec{b}_{St})_i \, (\Delta^{^{(2)}}
\vec{b^*})_j + (\vec{b}_{St})_j \, (\Delta^{^{(2)}} \vec{b}^*)_i \} $:
 $$
\hspace{-3mm}\Delta^{^{(2)}}\,{\cal{ A}}_{i\,j}^e(\tau)=  8 \,  g_{_{F'}}^2  \beta^2\,E_l \,\gamma \tau {\cal F}\, \left(
(\vec{B_t }\cdot \vec{E_t})
(\hat E_l \wedge \hat \epsilon_{ex})_i\, \, (\hat\epsilon_{ex})_j -(\vec{B_t}\cdot \,\hat\epsilon_{ex} ) (\hat E_l \wedge \hat
\epsilon_{ex})_i\,(\vec E_t)_j +( i
\leftrightarrow j )\right) .  
 $$
Note that the first term in the above  expression of  $\Delta^{^{(2)}}\,{\cal{ A}}_{i\,j}^e $ is proportional to
 $\Delta_{pv}\,{\cal{ A}}_{i\,j}^e$ (see Eq.(\ref{alinpv}) ) and, as a consequence, is expected 
 to  lead to a contribution to $ S_X-S_Y$  simulating the
true PV contribution,  if  $  \vec{E_t}\cdot \vec{B}_t  $ is even under the $ \vec{E_l} $ reversal.  
  
To get  the final  correction to the atomic imbalance   $ D^{at} $ we perform  the tensor 
  contraction of $ \Delta^{^{(2)}}\,{\cal{ A}}_{i\,j}^e(\tau) $ with the differential detector  tensor ${\cal{ D}}_{i\,j}$:
\begin{eqnarray}
 \Delta^{^{(2)}} D_{at}( \vec{E}_t, \vec{B}_t) = K  \Delta^{^{(2)}}\,{\cal{ A}}_{i\,j}^e(\tau) \cdot {\cal{ D}}_{i\,j} = 8
\,  g_{_{F'}}^2 
\beta^2\,E_l\, \gamma \tau \, K {\cal F}  \times  \hspace{40mm}\nonumber \\ 
\hspace{-15mm}\left [\left( (\vec{B_t }\cdot \vec{E_t}) 
 {\cal S }_{chir} (\hat\epsilon_{X}, \hat E_l)-( \vec{B_t} \cdot  \hat\epsilon_{ex} ) ( \hat E_l \wedge \hat \epsilon_{ex}
\cdot
\hat\epsilon_{X}) (\vec{E_t}\cdot \hat\epsilon_{X}) \right ) - (X  \leftrightarrow Y) \right]\; ,
\end{eqnarray}
where  ${\cal S }_{chir} (\hat\epsilon_{X}, \hat E_l) = -{\cal S }_{chir} (\hat\epsilon_{Y}, \hat E_l) $ is the pseudo-scalar
given
 by the formula  (\ref{chir}).
It is quite clear, as announced, that the first term, although being a  physical scalar, due to the
 pseudo-scalar   T-even factor $ \gamma \, \tau \,(\hat{B_t }\cdot \hat{E_t})$  is a potential 
source of sytematic error for the PV signal. In contrast to the PV-like term, the second term contains an anisotropic part.
  It is convenient to  rewrite  $D^{at}$ under a form  where  ${\cal S }_{chir} (\hat\epsilon_{X}, \hat E_l)$
 appears as an overall factor. To do this, we use  the formulae (20) defining
 $\hat\epsilon_{X} $ and $\hat\epsilon_{Y} $
 and  the explicit form of ${\cal S }_{chir} (\hat\epsilon_{X}, \hat E_l)$  given in Eq.(\ref{chir}).
 Performing explicitely the antisymmetrization with respect to the exchange $X  \leftrightarrow Y$, one arrives at the 
  expression:
 
\begin{equation}
 \Delta^{^{(2)}} D^{at}( \vec{E}_t, \vec{B}_t) =    16 \,  g_{_{F'}}^2  \beta^2\,E_l \, \gamma \tau \, K {\cal F} \, {\cal S
}_{chir} (\hat\epsilon_{X}, \hat E_l)  
\left(  \vec{B_t }\cdot \vec{E_t}-( \vec{B_t} \cdot  \hat\epsilon_{ex} )
( \vec{E_t}\cdot \hat\epsilon_{ex} )\right ) \, ,
\label{delDat}
\end{equation}
which exhibits the presence of a signal anisotropy under the global rotation of the experiment. Note that the last
parenthesis can be split into the sum of a purely isotropic contribution and a purely anisotropic one:
\begin{equation}
  \frac{1}{2}\vec{B_t }\cdot \vec{E_t}- \left (( \vec{B_t} \cdot  \hat\epsilon_{ex} )
( \vec{E_t}\cdot \hat\epsilon_{ex} ) - \frac {1}{2} \vec{B_t }\cdot \vec{E_t} \right ) .
\label{geomdep}
\end{equation} 

 We can write explicit expressions for the variations of this geometrical factor when global rotations of
the experiment are performed.   The initial position of $\hat \epsilon_{ex} \parallel \hat y$ in the laboratory
frame will serve as a reference axis.  If we  define the angles: $
(\hat y\,, \hat \epsilon_{ex})= \phi$, $(\hat y\,, \hat E_t\, ) = \theta_{E_t} $ and $(\hat y\,, \hat B_t\, ) = \theta_{B_t}$, we
obtain:  
 \begin{equation} 
 \left(  \hat{B_t }\cdot \hat{E_t}-( \hat{B_t} \cdot  \hat\epsilon_{ex} )
( \hat{E_t}\cdot \hat\epsilon_{ex} )\right ) = \frac{1}{2}   \left ( \cos{(\theta_{B_t} - \theta_{E_t}) - \cos{ (2 \phi - \theta_{B_t} -
\theta_{E_t})}   } \right ) \;.
 \end{equation}
We note that this expression is independent of the direction chosen as origin of angular coordinates, as it should be. 
Furthermore, the purely anisotropic contribution  has a dependence on the
rotation angle $\phi$ which only involves Fourier components at the frequency $2 \phi$. Consequently, four
measurements corresponding to $\hat
\epsilon_{ex} \parallel \hat y, \, \hat X,
\,  \hat x, \,$ and $ \hat Y$ are sufficient for us to obtain full information about the effect induced by the spurious fields
$\vec E_t,$ and $\vec B_t$. 
We notice that  $\Delta^{(2)} D^{at}(\vec E_t,\vec B_t)$ cancels when $\vec E_t $ is parallel to
$\hat \epsilon_{ex}$, as expected.

  This atomic imbalance, like the PV one,  is to be compared in magnitude to  that of the calibration
tilt angle (Eq.
\ref{calimb}):
\begin{eqnarray}
\theta_{cal} \frac{\Delta^{^{(2)}} D^{at} (\vec E_t,\vec B_t)}{D^{at}(\theta_{cal})} &=&  G_{EB}  
\left ( \cos{(2 \phi -\theta_{B_t} - \theta_{E_t})} - \cos{(\theta_{B_t} - \theta_{E_t})}  \right ) \;, \nonumber \\
{\rm where} \hspace{25mm} G_{_{EB}} &=&  \omega_{_{F'}} \tau \; \frac{E_t }{ E_l } \, (\hat z \cdot \hat E_l ) = \gamma
B_t \tau \; \frac{E_t }{ E_l }\, (\hat z \cdot \hat E_l) \; . 
\label{EBimb}
\end{eqnarray}
 
Like the PV one (Eq.(\ref{Datpvcal})), this ratio is odd under $\vec E_l $ reversal, if the quantity 
$\vec E_t \cdot \vec B_t$ is even.
   Therefore, we have shown the possible existence of a 
magnetoelectric optical effect which, in a given
$(\hat \epsilon_{ex},
\hat \epsilon_{pr}) $ configuration, can contribute to the APV quantity, the $E_l$-odd, calibrated linear dichroism $G$,
defined by Eq. (\ref {dichro}).  
\subsection{Properties of the isotropic and anisotropic contributions}
 The signal corresponding to expression (\ref{delDat}) has the symmetry properties
expected for a signal of electromagnetic origin. It is invariant under space reflexion.
 The isotropic contribution only differs from the APV signal by the pseudoscalar factor  $\vec E_t \cdot \vec B_t$ which 
cancels out only if the two transverse fields are orthogonal.  It is also invariant under time reversal since 
the precession angle $\omega \tau$   and $\vec B_t$ are both T-odd quantities.  
 This isotropic contribution mimics the APV signal only if its
behaviour in $E_l$-field reversal, like that of the APV signal is odd,  that is to say if $(\vec E_t \cdot
\vec B_t)$ is itself even under this reversal: {\it i.e.}~both fields are even or both odd. Hence there are two different
contributions. When we perform successive simultaneous rotations of $\hat \epsilon_{ex}$ and
$\hat \epsilon_{pr}$ by increments of 45$^{\circ}$, $\hat \epsilon_{ex} = \hat y, \hat X, \hat x, \hat Y  \; (\phi =
0^{\circ}, 45^{\circ}, 90^{\circ}, 135^{\circ}) $,  the transverse  fields remain fixed. Measuring the four calibrated linear
dichroism signals
$G_y, G_Y, G_x, G_X
$, actually provides us with two independent evaluations of the isotropic part: 
\begin{eqnarray}
 {\rm S_{_{xy}}} & = & \frac{1}{2} (G_x + G_y) = \theta^{pv} + \overline {G (\vec E_t,\vec B_t)} \;, \nonumber \\
 {\rm S_{_{XY}}} & = & \frac{1}{2} (G_X + G_Y) = \theta^{pv} + \overline {G (\vec E_t,\vec B_t)} \; ,
\label{sum}
\end{eqnarray}
where 
\begin{equation}
\overline {G (\vec E_t,\vec B_t)} = - G_{_{EB}}^{^{+}} \cos{(\theta
_{B_t^{+}}-\theta_{E_t^{+}})} -  G_{_{EB}}^{^{-}} \cos{(\theta
_{B_t^{-}}-\theta_{E_t^{-}})} .
\end{equation}
We have used the $+$ and $-$ superscripts to distinguish the case where $\vec E_t $ and $\vec B_t$ are both even or both
odd
 in $\vec E_l$ reversal.
The important point is that we expect - and we observe \cite{bou02} - equality to within noise between ${\rm S_{_{xy}}}$
and ${\rm S_{_{XY}}}$ \footnote{Let us note that in the case where $\vec E_t $ would exhibit a $\cos^2{(\phi)}$
modulation, though considered unlikely (\S~3.6) then the linear dichroism signals $G_y, G_Y, G_x, G_X $ would acquire a
$\cos{(4 \phi)}$ modulation. It can be predicted that this would lead to a difference between ${\rm S_{_{xy}}}$ and ${\rm
S_{_{XY}}}$, which is not observed in our measurements. $ {\rm D_{_{xy}}}$ and ${\rm D_{_{XY}}}$ would not be
modified.}.    
 On  the other hand, the signal anisotropy can be evaluated from the differences ${\rm D_{xy}}\equiv \frac{1}{2}(G_x -
G_y)$ and 
${\rm D_{_{XY}}} \equiv \frac{1}{2}(G_X - G_Y)$:
  \begin{eqnarray} 
 {\rm D_{_{xy}}} &=& -  G_{_{EB}}^{^{+}} \;  \cos {  (\theta_{E_t^{+}} + \theta_{B_t^{+}}) }  -  G_{_{EB}}^{^{-}} \;  \cos { 
(\theta_{E_t^{-}} +
\theta_{B_t^{-}}) } \;,  \nonumber \\   
{\rm D_{_{XY}}} &=& - G_{_{EB}}^{^{+}} \;  \sin {  (\theta_{E_t^{^{+}}} + \theta_{B_t^{+}})} - G_{_{EB}}^{^{-}} \;  \sin { 
(\theta_{E_t^{-}} +
\theta_{B_t^{-}})}\;\;.
\label{diff}
\end{eqnarray}
 In order to obtain $\overline {G (\vec E_t,\vec B_t)}$ we need additional information we can actually extract from
complementary measurements performed in presence of an auxiliary magnetic field. Measurements similar to the PV
ones made in such a field of known direction and magnitude (~2 G) 
 yield the directions and magnitudes of $\vec E_t^{+}$ and $\vec E_t^{-}$, the transverse field contributions even and odd in
$\vec E_l$ reversal.  However, as we shall see in the section 5, the situation is somewhat complicated by the
presence of symmetry breaking effects of another kind which lead to purely  anisotropic contributions.
\subsection{Comment on the $\vec E_t \cdot \vec B_t$ induced magnetoelectric dichroism}
 We have just demonstrated theoretically the existence in our
atomic system of a linear dichroism generated by two parallel, transverse fields  $\vec E_t$ and $\vec B_t$. Its magnitude
is proportional to
$\vec E_t \cdot \vec B_t$ and its axes are oriented at $\pm 45^{\circ}$ to the field  direction.  Our observations confirm
the expected magnitude (see \S.~6 for a more detailed  discussion). We have actually exploited it thoroughly for controlling
the stray $E_t$-fields generated in Cs filled alumina cells by electric charges produced by photoionization \cite{gue02}.

Such an effect supported by symmetry arguments was predicted long ago by Jones \cite{jon48}.  
Despite searches in molecular liquids, it remained unobserved until recently
\cite{rik01}, when the samples were placed under extreme experimental conditions: a $B_t$ field of 15 T and an $E_t$ field
of 1.75 kV/cm. Therefore it is worth underlining the differences between our present experimental conditions and those
which correspond to the molecular liquid samples used for this kind of observation: number densities about 
$10^6$ times lower, magnetic fields  
$10^4$ to $10^7$ times smaller, and a transverse
$\vec E_t$  field about one hundredth of the size of the longitudinal
$\vec E_l$ field of comparable magnitude which provides the vapor with axial symmetry. Above all one should bear in
mind that in the present case we have a {\it dilute  atomic} sample in a particularly simple atomic state. Consequently, the
effect of the magnetoelectric  dichroism observed on the probe beam can be given a simple, intuitive and quantitative
interpretation: the Larmor precession of the atomic alignment generated in the excited state by interference between the
Stark-induced dipoles involving both the longitudinal and transverse components of the electric field.  But this mechanism is
not unique. For other illustrations which lend themselves to detailed calculations also in the context of atomic vapors excited
via a highly forbidden transition,  see for instance the forthcoming section 
4.5 and reference \cite{bud03}. 

\subsection{Combined effect of an $\vec E_t$ field and a pump-probe misalignment} 
 We have  shown in subsection 4.1 that the alignment  associated with a small  transverse electric field  $ \vec E_t  $
leads to an atomic imbalance in presence of a transverse magnetic field $ \vec B_t  $. (The case of an $ E_l$-odd $\vec
E_t$ field accounts for a misalignment between the applied $\vec E_l$ field and the excitation beam direction $\hat
k_{ex}$).  We now want to  show that a misalignement of the probe beam  with respect to the excitation beam leads to the
same effect.  This deviation  from the ideal configuration can be described by applying to the polarized probe beam an
infinitesimal rotation by a  small angle
$\delta\alpha$  about a unit vector $\hat n $ normal to the excitation photon momentum $
\vec{k}_{ex}$ such that $ \delta \alpha \, \hat n =
\hat k_{ex}
\wedge \hat k_{pr}$.  

 Such an infinitesimal rotation of the probe beam leads to a correction $
{\Delta}_{mis}^{^{(1)}} {\cal D }_{i\,j} $ of the detection tensor (Eq. (17)) which is given by:    
 \begin{equation}
  {\Delta}_{mis}^{^{(1)}}{\cal D }_{i\,j}=  \delta\alpha \left( (\hat n \wedge \hat\epsilon_X)_i \, ( \hat\epsilon_X)_j+ 
( \hat\epsilon_X)_i \,(\hat n \wedge \hat\epsilon_X)_j - (X  \rightarrow  Y) \right) . 
\label{delmisDet}
\end{equation}
By combining the effect of the probe beam tilt with the  transverse  electric field  $ \hat E_t  $ correction to the alignment 
tensor $ {\Delta}^{^{(1)}} {\cal A }_{i\,j}^e\propto b_i \,   {\Delta}^{^{(1)}}_{E_t} b_j +(i  \leftrightarrow j  ) $, we predict a
second-order correction to  the atomic  imbalance $ D^{at} $:
$$   {\Delta}^{^{(2)}}\, D^{at}(E_t, mis) = K\,   {\Delta}^{^{(1)}} {\cal A }_{i\,j}^e(0)  \cdot  {\Delta}_{mis}^{^{(1 )}}{\cal D
}_{i\,j} .
$$ Performing explicitely the tensor contraction  together with the  antisymetrisation with respect to $X \leftrightarrow  Y$, 
we obtain the final  expressions:
\begin{eqnarray}   
{\Delta}^{^{(2)}} \, D^{at}(E_t, mis) = -16   \,  g_{_{F'}}^2  \beta^2\,E_l \, \delta \alpha \, E_t \,K \,{\cal F}\, 
 {\cal S }_{chir}
(\hat\epsilon_{X}, \hat E_l)  
\left(  \hat{n}\cdot \hat{E_t}-( \hat{n} \cdot  \hat\epsilon_{ex} )
( \hat{E_t}\cdot \hat\epsilon_{ex} )\right ) \, , \nonumber \\
\theta_{cal} \,  \frac{\Delta^{^{(2)}} \, D^{at}(E_t ,mis)}{ D^{at}(\theta_{cal})} = G_{_{E\,\delta \alpha}} 
\left(  \hat{n}\cdot \hat{E_t}-( \hat{n} \cdot  \hat\epsilon_{ex} )
( \hat{E_t}\cdot \hat\epsilon_{ex} )\right ),\; 
 {\rm with} \;  G_{_{E \,\delta \alpha}} =  \delta \alpha  \frac{E_t}{E_l } \, (\hat z \cdot \hat E_l )\;.
\label{misalDat}
\end{eqnarray}  
 Thus only an $\vec E_l {\rm -}$even transverse field, $\vec E_t^{+}$, can give rise to a harmful $\vec E_l {\rm -}$odd
imbalance. It is also interesting to note  the close similarity of the above result with that induced by $E_t,  B_t$ transverse
fields Eq. (\ref{delDat})  in \S~4.1,  with just a single substitution performed on the P-even T-even
rotational invariant:  
\begin {equation}
  \omega_{_{F'}} \hat B_t\hspace{3mm} \longrightarrow \hspace{3mm}  - \, \delta
\alpha
\,   \hat n  \;. 
\end{equation}

It is actually not surprising that the Larmor  precession and a small rotation of the probe polarized beam lead to similar
effects. Both effects arise from the same correction to the second-rank tensor
$\Delta^{^{(1)}}{\cal A}_{i,j}^e$, via either a small rotation of this tensor itself about $\hat B_t$ in the former case, or a small
rotation of the detection tensor about $\hat n$ in the latter case.
 The above  exact correspondance, and in particular the change  of sign,   can 
be understood  from rather simple rotational invariance considerations \footnote{   Let us first  illustrate this point for  
a simpler configuration:  $ \vec{a} $ and $ \vec{b} $  are assumed to be 
 two physical  3-dimensional vectors and $ R ( \hat n,\phi) $ a rotation  acting  only  upon the vector $ \vec{b} $. 
Let us  assume further that  the scalar product 
 $\vec{a}\cdot (R ( \hat n,\phi) \,\vec{b })$ is associated with some physical observable; it is invariant under  
any global rotation $ R' $  acting on the two vectors $ \vec{a} $ and $ \vec{b} $.  By choosing $ R'$  to be the inverse of
$R$: 
$ R'= R^{-1} ( \hat n,\phi) = R ( \hat n,-\phi) $, we get immediately the identity:
 $$ \vec{a}\cdot( R ( \hat n,\phi) \,\vec{b })= R'\vec{a}\cdot  (R'\,R ( \hat n,\phi) \,\vec{b })=
( R^{-1} ( \hat n,\phi) \vec{a})\cdot \vec{b}\;.  $$  
The physical content of the above equation can be stated as follows: 
{\it For  a physical measurement involving the scalar product of two vectors, the effect of a 
rotation $R$ acting upon the vector on the r.h.s. is identical to that of the inverse rotation  $R^{-1} $ acting  upon the vector on
the l.h.s.}. 

What  is involved here  is actually the same idendity, but applied  to  the contraction 
 of two second-order tensors, $A_{i\,j}\cdot B_{i\,j} $ . It is not difficult to generalize. We can 
begin by  considering the case of
two  factorized tensors: $  A_{i\,j}= a_i\,b_j $ and $  B_{i\,j}= c_i\,d_j  $. The contraction  of them reads: 
$A_{i\,j}\cdot B_{i\,j}=  (\vec{a}\cdot  \vec{c})\, (\vec{b}\cdot  \vec{d})$. Let us rotate the r.h.s tensor:
 $ B_{i\,j} \rightarrow B_{i\,j }^R= (R\, \vec{c})_i \,(R\, \vec{d})_j$. We  get immediately: 
$A_{i\,j}\cdot B_{i\,j}^R =(\vec{a}\cdot R\,\vec{c} ) (\vec{b}\cdot  R  \,\vec{d}) =
(R^{-1}\vec{a}\cdot \vec{c} ) (R^{-1}\vec{b}\cdot \vec{d})=A_{i\,j}^{R^{-1} }\cdot B_{i\,j}$. Then, the property can be extended to the general
case  by noting that 
 an arbitrary  second-order tensor  $A_{i\,j}$  can always be written as a linear combination of the nine factorized tensors
constructed from   three independent vectors: $ \vec{a}_1,  \vec{a}_2 , \vec{a}_3 $.  }.
\subsection{Other mechanisms involving the $M_1$ amplitude}
As already noted in \S~2.2.4, for there to be a systematic effect involving the $M_1$ amplitude there must also be a a
transverse electric field. We also stressed that such a field alone is not enough. We now show how combining
both electric and magnetic transverse fields can generate a systematic effect. Since the typical value of $\vert M_1/ \beta
E_l \vert $ in our experiments is $\simeq 2 \times 10^{-2} $, it looks like we could consider  an $ M_1$-Stark-induced
contribution as a third order effect. In reality, because the presence of a
$\vec B_t$ field makes slight mixing of the different hyperfine states, the $M_1$-Stark interference effect is
enhanced. In fact, given that the hyperfine substates are no longer  pure F states, the scalar Stark amplitude 
$\alpha \vec E_l \cdot \hat \epsilon_{ex}$ can connect S-states having different hyperfine quantum numbers. 
As a result, the size of the perturbation is
reinforced by a factor  $ \vert \alpha /\beta\vert  \simeq 10$. This is why we now want to focus our discussion on
this particular contribution.

The matrix element of the scalar Stark-induced amplitude between the 
$6S,F$ and the $7S, F'$ states with $F' \not= F$ is calculated using first
order perturbation theory \footnote{We deliberately omit here the magnetic perturbation of the $6S,F$ state which 
has a larger hyperfine structure and leads to an effect about 5 
times smaller.}. It  is convenient to write the result in terms of a  
correction contribution to the effective dipole $\vec d^{eff} $(see Eq. (1)):
 $$
\Delta^{^{(2)}}_{\alpha E_t, B} \vec d^{eff} = -\alpha \vec E_t  (F' - F) 
\frac{\gamma_S B}{\Delta W_{7S}} P_{_{F'}} \vec \sigma \cdot \hat B 
P_{_F} \;,  
 $$
where $\gamma_S  (=  \gamma/g_{_F})$ is the gyromagnetic factor for the electron spin. We write 
the resulting  correction
to the complex  vector $ \vec{b} $ which   appears in the effective transition matrix $ T_{eff} $  (see Eq. (5)) as:
 $$ \Delta^{^{(2)}}_{\alpha E_t, B} \vec b\ = \alpha (\vec E_t \cdot \hat \epsilon_{ex})  (F' - F) 
  \frac{\gamma_S
  B_t}{\Delta W_{7S}} P_{_{F'}}  \hat B_t P_{_F}. 
$$
We insert the above result in the general expression of the excited state 
alignment, Eq. (13), to get a new   $ M_1 $ amplitude contribution 
$ \propto {\rm Re }\{  (b_{ M_1 })_i \Delta^{^{(2)}}  
b_j +(  i \leftrightarrow j )\} , $
where  $\vec{ b}_{ M_1 } = -M_1 \hat k \wedge \hat \epsilon_{ex}$.
We now have all we need to write the alignment of the
excited state resulting from the interference between the 
$M_1$ and the hyperfine-mixing  scalar Stark amplitudes:
\begin{equation}
  \Delta^{^{(3)}}_{(M_1 \, \alpha E_t,  B)} {\cal A}_{i\, j} =   4 g_{_{F'}}^2  \, {\cal F} 
(F' - F) \frac{\gamma_S B}{\Delta W_{7S}} \left ( M_1 \alpha \vec E_t 
\cdot \hat \epsilon_{ex} \right )
  \left ( (\hat B_t)_i  (\hat k \wedge \hat \epsilon_{ex} )_j  + ( i  \leftrightarrow j ) \right ) \,.
\end{equation}

Contracting  this tensor  with the detection tensor ${\cal D}_{i\,j} $
we get the following contribution  to the atomic  imbalance:
\begin{eqnarray}
  \Delta^{^{(3)}}   D^{at}(M_1, \alpha E_t,\,B_t)& = &  16 g_{_{F'}}^2  \, {\cal F} K (F' - F ) \frac{\gamma_S
}{\Delta  W_{7S}}  M_1 \alpha \, {\cal S}_{chir}(\hat \epsilon_X, \hat E_l) 
(\vec E_t \cdot \hat \epsilon_{ex})(\vec B_t \cdot \hat \epsilon_{ex}) \, (\hat k \cdot \hat E_l )   \nonumber  \\
&=&  16 g_{_{F'}}^2  \, {\cal F} K\,(F' -F ) \frac{\gamma_S }{\Delta W_{7S}}  M_1 \,
\alpha \,{\cal S}_{chir}(\hat
\epsilon_{X}, \hat E_l)   \, (\hat k \cdot \hat E_l )\times \nonumber \\
 & &  \left [ \frac{1}{2} \vec  E_t \cdot \vec B_t  + \left ( (\vec B_t \cdot \hat\epsilon_{ex})
  (\vec E_t \cdot \hat \epsilon_{ex}) - \frac {1}{2}\vec E_t \cdot 
\vec  B_t  \right )\right].
\label{interfalphaM1}
\end{eqnarray}
  The last factor in the r.h.s of the above  equation  has been split into its isotropic and its purely 
anisotropic parts.
It is interesting to compare it with that obtained previously for the 
$\vec E_t\cdot \vec B_t$ effect, Eq. (\ref{geomdep}): the only  difference lies in  a reversal 
of the {\it relative sign} of these two contributions. 
 To satisfy the $\vec E_l {\rm -}$odd behavior, {\it  here, in contrast to the previous case, the scalar product 
$\vec E_t \cdot \vec B_t$  has to be odd under $\vec E_l $ reversal}. This is why it has been considered here, even 
though it would be about 20 times smaller were the $\vec E_l$-odd and  $\vec E_l$-even contributions to $ \vec E_t
\cdot \vec B_t$ of equal magnitude.  

One may note also the presence of the additional factor $(\hat k \cdot \hat E_l)$  in the
above expression (\ref{interfalphaM1}): it has appeared in a natural way via the
contribution of the $M_1$ amplitude and, combined with the factor
$(\vec B_t \cdot
\hat \epsilon_{ex})$, it ensures T-reversal invariance of the result   
in which, by contrast to
the  Larmor precession dependent contribution (Eq. (\ref{delDat})), the time no longer appears explicitly.   
  
Although we are not going to present here  the details of the corresponding  calculation, 
 we would like to mention that the atomic imbalance associated to  
the vector Stark amplitude, instead of the
scalar one,  involves this time the   same angular  dependence   as that given  by Eq. (\ref{geomdep}).
 Note that this contribution  is suppressed by a factor $ \vert \beta/\alpha \vert =1/10 $
 with respect to that given by Eq. (\ref{interfalphaM1}).

 For completeness, let us quote the result deduced from Eq.(\ref{interfalphaM1}) after normalization by the 
calibration imbalance (Eq. (\ref{calimb})):
\begin{equation}
 \theta_{cal} \frac {\Delta^{^{(3)}} D^{at} (M_1, E_t,  B_t)}{D^{at}(\theta_{cal})}  =  - (F' - F) \frac{\gamma_S B_t}{\Delta
W_{7S}}  
\frac{M_1 \alpha E_t}{(\beta E_l)^2} 
(\hat E_t \cdot \hat \epsilon_{ex})(\hat B_t \cdot \hat \epsilon_{ex}) (\hat k \cdot \hat E_l)\,.   
\end{equation}
 
\section{Symmetry breaking by two transverse magnetic fields, $\vec E_l$-odd and $\vec E_l$-even }
Let us suppose that two transverse magnetic fields, $\vec B_t^{+}$ and $\vec B_t ^{-}$, one
even and the other odd under $\vec E_l$-reversal are now the source of cylindrical symmetry breaking. Since the
dominant Stark alignment is $\vec E_l$-even, those two fields having opposite behavior under $\vec E_l$-reversal can
introduce an
$\vec E_l$-odd character in the effect arising from perturbation to second order. Different possible mechanisms can enter
into play, so we shall first consider one of them in detail. Afterwards, we shall present a generalization of the main features
of the results.
\subsection{Larmor precession of the Stark alignment about a $\vec B_t$ field changing its direction under 
$\vec E_l$ reversal }
We consider first the effect of Larmor precession. We shall show that the dominant Stark alignment 
(Eq. (\ref{stalin})) precessing in a transverse magnetic field acquires a detectable component of second-order in the
infinitesimal precession angle $\omega_{_{F'}} \tau = \gamma \tau B_t$. The variation of $\vec b_{St}$ induced by the
Larmor precession can be written:
\begin{equation}
\Delta_L^{^{(1)}} \vec b_{St} = i \beta \, \gamma \,\tau \, \vec{B}_t \wedge (\vec{E}_l \wedge \hat
\epsilon_{ex} )= i \beta \, \omega_{_{F'}} \,\tau \left ((\hat B_t \cdot \hat\epsilon_{ex})\vec{E_l} - (\hat {B}_t \cdot
\vec{E_l})
\hat \epsilon_{ex}
 \right) .
\label{delLSt}  
\end{equation}
 Since $\vec B_t \cdot  \vec{E_l} = 0 $, the second term cancels, and since $\vec E_l$ is along $\hat k_{pr}$, it can be easily
verified that the second order correction to the alignment of the type $(\Delta_L^{^{(1)}} \vec b)_i (\Delta_L^{^{(1)}} \vec
b^*)_j + i
\leftrightarrow j $  cannot contribute to a polarimeter imbalance. Even so, a possible misalignment between
the pump and the probe beams combining to the first order Larmor correction can contribute, a situation that will be
considered later on (\S~5.3). 
 
 For the moment, we calculate the second-order correction to $\vec b$
 associated with an infinitesimal rotation angle $\phi= \omega_{_{F'}} \tau $ about $\hat B_t$ to second order in
$\phi$.
We can separate $\vec b$ into its components parallel and perpendicular to the rotation axis:
 $$\vec b = \vec b_{\parallel}+ \vec b_{\perp} =  (\vec b\cdot \hat B_t ) \; \hat B_t  + \left ( \vec b - (\vec b\cdot \hat B_t ) \;
\hat B_t\right ) \, . $$
 The transformed vector is then written:
 $${\cal R} \vec b = \vec b_{\parallel} + \vec b_{\perp}\cos{\phi} + \hat B_t \wedge \vec b_{\perp} \sin{\phi} \;,$$  
 which we expand to second order in $\phi $:
\begin{eqnarray}
{\cal R} \vec b & = & \vec b_{\parallel} + \vec b_{\perp} (1 - \frac{\phi^2}{2}) + (\hat B_t \wedge \vec b_{\perp}) \;
\phi +{\cal O}(\phi^3) \nonumber \\ 
&=&  \vec b + (\hat  B_t \wedge \vec b) \; \phi + \left ((\vec b \cdot \hat B_t) \hat
B_t - \vec b \right ) \frac{\phi^2}{2} +{\cal O}\phi ^3 .
\end{eqnarray}
In the present situation, we are only interested in the contribution which alters the direction of 
$\vec b_{St}$ along $ \hat h = \hat E_l \wedge \hat \epsilon_{ex} $ at $\tau = 0$:
\begin{equation}
\Delta_{L}^{^{(2)}}\vec b_{St}=  i \beta \, \frac{\omega_{_{F'}}^2 \,\tau^2}{2} \left( \hat E_l \wedge \hat \epsilon_{ex}
\cdot
\hat B_t   \right) \hat B_t  \;.
\end{equation}
 This gives rise to a second order correction to
${\cal A}_{ij}^{St}(\tau)$ obtained by replacing in Eq. (\ref{stalin})
${\rm Re}
\{b_i b_j ^*\}$ by $\left ( (\vec b_{St})_i ( \Delta^{^{2}}_L\vec b_{St})_j  + i \leftrightarrow j \right )$, which leads 
to the detectable alignment :
\begin{equation}
\Delta_{L}^{^{(2)}} {\cal A}_{ij}^e(\tau) =  2 \,  g_{_{F'}}^2  \beta^2\,E_l \, \omega_{_{F'}}^2 \tau^2 \, {\cal F} (\hat h
\cdot
\hat B_t)\left (\hat  h _i  \; (\hat {B_t}) _j  + i \leftrightarrow j\right ) \;. 
\end{equation}
 After contraction with the second-rank detection tensor, the last factor in parentheses can be written $ 4 \, {\cal S}_{chir}
(\hat \epsilon_X, \hat E_l)\, (\vec B_t  \cdot \hat \epsilon_{ex})  $.
This imbalance would not be $\vec E_l$-odd unless
$\vec B_t$ were the sum of two fields having opposite behaviour under
$\vec E_l$ reversal ($\vec B_t = \vec B_t^{+} + \vec B_t^{-}$), which means that $\vec B_t$ would change direction
when
$\vec E_l$ is reversed. We arrive at the final result for the
$\vec E_l$-odd imbalance induced by two transverse magnetic fields respectively odd and even: 
\begin{eqnarray}
\Delta_{L}^{^{(2)}} D^{at}(\vec B_t^{+}, \vec B_t^{-}) =  8 \,  g_{_{F'}}^2  \beta^2\,E_l^2 \,
\omega_{_{F^{\prime}}}^{+}\omega_{_{F^{\prime}}}^{-} \tau^2 
 K \, {\cal F}  \;{\cal S}_{chir}(\hat \epsilon_{X}, \hat E_l) \, \times  \nonumber \\
 \left ( (\hat
B_t^{+} \cdot \hat E_l \wedge \hat \epsilon_{ex})(\hat B_t^{-} \cdot \hat \epsilon_{ex}) +  \hat B_t ^{-}
\longleftrightarrow
\hat B_t^{+} \right ),
\label{DatBB} 
\end{eqnarray} 
 where $\omega_{_{F'}}^{+}= \gamma B_t^{+},$ and $\omega_{_{F'}}^{-}= \gamma B_t^{-}$. After calibration:
\begin{equation}
\theta_{cal} \frac{\Delta^{^{(2) }}_L D^{at}(\vec B_t^{+}, \vec B_t^{-})}{D^{at}(\theta_{cal})}= - 
\omega_{_{F^{\prime}}}^{+}\omega_{_{F^{\prime}}}^{-}\tau^2  \, \left ( (\hat B_t^{+}  
\cdot \hat  z \wedge \hat \epsilon_{ex} \,)(\hat B_t^{-} \cdot \hat \epsilon_{ex}) +  \hat B_t ^{-} \longleftrightarrow
\hat B_t^{+}  \right ).
\label{imbBB}
\end{equation}
When global rotations of the experiment are performed, the behavior of this signal is purely anisotropic. 

\subsection{Generalization of the result to second-order magnetic perturbations of the pump and probe transitions}      
Let us consider the perturbation of the probe transition by the transverse magnetic field $\vec B_t^{+} +
\vec B_t^{-}$ which is well known (see for instance \cite{bud02}) to be responsible for a linear dichroism quadratic
in this field, the so-called Voigt effect. Here we are interested in its $\vec E_l$-odd contribution $\propto B_t^{+} B_t^{-}$.
 In the evaluation of the second-rank tensor operator ${\cal O}_{ij}$ (Eq.~(\ref{seop})), the $\vec B_t $ field shows up in
two different ways:

 - it modifies the wavefunctions and thus the transition probabilities;

 - it shifts the energy levels. 
 
The magnitude of the perturbation is given by the ratio of the Zeeman frequency shift to the line width: 
$\omega_{_{F}} /\Delta \omega \sim  10^{-5}  $
per milligauss.  In the present context, this effect can be considered as a second-order modification of the detection tensor (see
Eq. \ref{dettensor}) by the transverse magnetic field.  We obtain this  $\Delta ^{(2)} {\cal D}_{i\, j}$ modification by
performing the contraction  of the two second-rank tensors, $\hat B_j \hat B_k $ which represents the perturbation  and the
unperturbed ${\cal D}_{i \, k}$ detection tensor:
$$ \Delta ^{^{(2)}}{\cal D}_{i \, j} \sim  \left (  \frac{\omega_{_{F'}}}{\Delta \omega} \right )^2 (\hat B_t)_i  (\hat B_t)_k
{\cal D}_{j
\, k}
\sim \left (  \frac {\omega_{_{F'}}}{\Delta \omega} \right )^2  \left ((\hat B_t)_i (\hat B_t \cdot \hat \epsilon_X)(\hat
\epsilon_X)_j - ( X
\leftrightarrow Y)\right ) .     
$$ 

When this detection tensor modification is contracted with the Stark alignment tensor, we get $ \Delta^{^{(2)}} D^{at} (\vec
B_t^{+},
\vec B_t^{-}) = \Delta ^{(2)}{\cal D}_{i \, j} \cdot {\cal A}_{i\, j}^{St} = 
 \left (\frac{\omega_{_{F'}}}{\Delta \omega} \right )^2 \left ( (\hat B_t \cdot \hat \epsilon_X)(\hat B_t
\cdot \hat h)(\hat h \cdot \hat \epsilon_X) - (X \leftrightarrow Y) \right)$  
This yields a second-order correction to the
atomic polarimeter imbalance:
\begin{equation}
  \theta_{cal} \frac{\Delta^{^{(2)}} D^{at} (\vec B_t^{+}, \vec B_t^{-})}{D^{at}(\theta_{cal})} \sim  
\left (  \frac{\omega_{_{F^{\prime}}}^{+}\omega_{_{F^{\prime}}}^{-}}{\Delta \omega^2} 
\right ) \left ( (\hat B_t^{+} \cdot \hat \epsilon_{ex}) ( \hat B_t^{-} \cdot \hat z \wedge \hat \epsilon_{ex}) 
 + ( + \leftrightarrow  - ) \right )
\, ,      
\end{equation}
where the expression into  parentheses is actually identical to that appearing in Eq. (\ref{imbBB}). Note that the order of  
magnitude is also relatively close since $\Delta \omega \, \tau $ is not very different from unity. 

The treatment given above can be applied without any difficulty to the second-order magnetic perturbation of the pump transition, leading to a very
similar result. We conclude that the magnitude and the structure of the result is valid for any kind of second-order
magnetic perturbation of the atomic system. 
\subsection{Combined effect of a misalignment and a magnetic perturbation}
As mentioned previously,  it is possible to generate
a $\vec E_l$-odd polarimeter imbalance by combining a magnetic perturbation and a pump-probe misalignment. This can
be viewed as resulting from the contraction of the first-order perturbation, 
$\Delta_{mis}^{^{(1 )}}{\cal D}_{i \, j}$ (Eq. (\ref{delmisDet})) to the detector tensor and the first-order Larmor perturbation 
$\Delta_L^{^{(1)}}{\cal A}_{i \, j}^{St}(\tau)= - 4 g_{_{F'}}^2 \left ((\Delta_L^{^{(1)}} \vec b_{St})_i (\vec b_{St})_j  + ( i
\leftrightarrow j ) \right )$  to the Stark alignment ${\cal A}_{i \, j}^{St}(\tau)$  (Eq. (\ref{delLSt})), we calculate the $\vec
E_l$-odd atomic imbalance:

\begin{eqnarray}
\Delta_{misL}^{^{(2)}} D^{at} = \Delta_{mis}^{(1)} {\cal D}_{i \, j} \cdot  \Delta_L^{^{(1)}} {\cal A}_{i \, j}^{St}(\tau) 
\hspace{80mm}\nonumber
\\
 =  - 8 \,  g_{_{F'}}^2  \beta^2\,E_l^2 \, \omega_{_{F'}} \tau \delta \alpha \,  
 K \, {\cal F}  \;{\cal S}_{chir}(\hat \epsilon_{X}, \hat E_l) \, \left ( ( \hat n \cdot \hat E_l \wedge \hat \epsilon_{ex})(\hat
B_t^{-}
\cdot
\hat
\epsilon_{ex}) + (\hat B_t^{-}
\longleftrightarrow \hat n) \right ),
\end{eqnarray}
and the calibrated imbalance:
\begin{equation}
\theta_{cal} \frac{\Delta_{misL}^{^{(2)}} D^{at}(\vec B_t^{+}, \vec B_t^{-})}{D^{at}(\theta_{cal})}=   \omega_{_{F'}} \tau
\delta
\alpha 
 \, \left ( (\hat n  
\cdot \hat z \wedge \hat \epsilon_{ex} \,)(\hat B_t^{-} \cdot \hat \epsilon_{ex}) +(\hat B_t^{-} \longleftrightarrow \hat
n)
\right ).
\end{equation}
We note again the expected correspondance between this result and that relative to Larmor precession treated to second order 
(Eq. (\ref{imbBB})), when one performs the substitution:
$$ \omega_{_{F'}} \,   \hat B_t^{+}   \hspace{3mm} \longrightarrow \hspace{3mm}  - \, \delta \alpha
\,   \hat n  \;. $$
 
\subsection{Absence of isotropic contribution. Anisotropy properties}
For all the effects considered in this section 5, the most important property already mentioned, is the {\it absence of 
an isotropic contribution}. In the quantities
${\rm S_{xy}} = \frac{1}{2}( G_x+ G_y)$ and
${\rm S_{XY}}= \frac{1}{2}(G_X + G_Y)$, the
 spurious $\vec B_t ^{+}\, \vec B_t^{-}$ effect cancels out.
Therefore, combining results obtained in two different input $\hat \epsilon_{ex}$ polarization
configurations is sufficient to eliminate it as a source of systematics. 
However, some inconvenience remains as a result of the associated anisotropy. 

 If we introduce the angles $\theta_{B_t}^{+}= (\hat y_0, \vec B_t^{+})$ and
$\theta_{B_t}^{-}= (\hat y_0, \vec B_t^{-})$,  this anisotropy is described by the angular dependence: 
$$ \left (\sin(\phi - \theta_{B_t}^{-}) \cos(\phi -\theta_{B_t}^{+})  +  \theta_{B_t}^{-} \longleftrightarrow
\theta_{B_t}^{+}\right ) = \sin{(2 \phi - (\theta_{B_t}^{+} + \theta_{B_t}^{-} ))} .
$$
 This is similar to the anisotropy caused by the $(\vec E_t \cdot \vec B_t)^{+}$ effect, but with different, and {\it a priori} not
simply related, anisotropy direction. Both effects can be present simultaneously. This is why 
we cannot extract $\overline {G (\vec E_t,\vec B_t)}$ from the measured anisotropy. 
In order to obtain more information we shall have to rely on another property to be discussed in \S ~6.3. 

\section{ Isotropy tests}
\subsection{ Symmetry breaking effects and ways to reduce them}
 We have listed in Table 1 the various candidates to systematic effects arising from cylindrical symmetry breaking 
defects. There are two main classes of such effects. Both have their magnitude characterized by a
2$\phi$-frequency modulation when the experiment is globally rotated. This implies that measurements in only four  
configurations obtained by  successive rotations of 45$^{\circ} $ are necessary to obtain full information.  Effects of the first
class are dangerous since an isotropic contribution remains after our averaging the linear dichroism signal over those four
configurations and represents a systematic effect superimposed on the APV signal. However, it is accompanied by an
anisotropic contribution of the same order of magnitude which can indicate  its presence. In order to reduce its magnitude
one may proceed by reducing the spurious $\vec E_t$ and
$\vec B_t$ fields as well as the pump-probe beam misalignment. Effects of the second class provide contributions which cancel
out when averaged over the four input polarization  configurations. They look harmless but in fact 
complicate the interpretation of the anisotropy, when present. This is why a correlation test is particularly
welcome based on statistical data analysis
which may establish  absence of any significant link between the anisotropic and isotropic contributions  (see \S~6.2).\begin{table}
\caption {Summary of the various signals contributing to the polarimeter imbalance, with the amplitudes involved in the
excitation process (1st column), the experimental defects involved (2nd column) and the angular dependence under global
rotation of the experiment (3rd column). The symmetry-breaking defects are represented by a pair of dimensionless vector
fields of small magnitude.  From the product of the quantities appearing in columns 2 and 3 of each line, we obtain the
atomic imbalance of the polarimeter (normalized by the calibration imbalance) which has to be compared with 
$\theta^{pv}$ to obtain the fractional importance of the systematic effect.  }

~~~~~~~~~~~~~

{\bf Origin} \hspace{70mm}{\bf  Angular Dependence} ~~ 

~~~~~~~~~~~~~~~~~~~~~~~~~~~~
 
~~~~~~~~

{\bf APV Signal}

~~~~~

$\beta E_l  E_1^{pv}\hspace{17mm} -\frac{{\rm Im} E_1^{pv}}{\beta E_l} = \theta^{pv}  \hspace {45mm}$ None 

~~~~~~~~

{\bf Possible Systematics: }

~~~~~~

interference  \hspace{8mm} defects involved\hspace{11mm} {\bf Class 1:  Effects with isotropic
contribution}

~~~~~~~~~~

$\beta E_l \beta E_t \hspace{18mm} \frac{E_t}{E_l} \hat E_t , \, \omega_F \tau \hat B_t \hspace{16mm}    \left (\hat
E_t^{+} \cdot
\hat B_t^{+} - (\hat B_t^{+}\cdot \hat \epsilon_{ex})(\hat E_t^{+} \cdot \hat \epsilon_{ex})   \right ) + (+  
\longrightarrow  - )
$

~~~~~~~~~~

$\beta E_l \beta E_t \hspace{18mm} \frac{E_t}{E_l} \hat E_t, \, \delta \alpha \,\hat n\hspace{35mm}   \left
(\hat E_t^{+} \cdot \hat n - (\hat n \cdot \hat \epsilon_{ex})(\hat E_t^{+} \cdot \hat\epsilon_{ex})   \right )$

~~~~~~~~~~

$M_1 \alpha E_t  \hspace{14mm}  \frac{M_1 \alpha E_t}{\beta^2 E_l^2} \hat E_t , \, \frac{\gamma_S B_t}{\Delta W_{7S}
}\hat B_t\; 
\hspace{23mm}  
\;(\hat B_t^{+}\cdot \hat \epsilon_{ex})(\hat E_t^{-} \cdot \hat \epsilon_{ex})+ (+ \longleftrightarrow - )$

~~~~~~~~~~~~~

\hspace{59mm} {\bf Class 2:  Effects without isotropic contribution} 
 
~~~~~~~~~~~~~~
 
$\beta^2 E_l^2 \hspace{18mm}\omega_{_{F'}}^{+}  \tau \hat B_t^{+}, \omega_{_{F'}}^{-}  \tau \hat B_t^{-} \hspace{20mm} -
(\hat B_t^{+} \wedge
\hat z \cdot \hat \epsilon_{ex})(\hat B_t^{-} \cdot \hat \epsilon_{ex}) + ( - \longleftrightarrow  + )  
$

~~~~~~~~~~~~~~

$\beta^2 E_l^2 \hspace{20mm}\omega_{_{F'}}^{-}  \tau  \hat B_t^{-},\delta \alpha \, \hat n\hspace{25mm}  ~~(\hat n \wedge
\hat z \cdot \hat \epsilon_{ex})(\hat B_t^{-} \cdot \hat \epsilon_{ex}) + ( \hat B_t ^{-}\longleftrightarrow \hat n)    
$

~~~~~~~~~~~~~~~~

$ \beta^2 E_l^2   \hspace{19mm} \frac{\omega_{_{F'}}^{+}}{\Delta \omega } \hat B_t^{+},  \frac{\omega_{_{F'}}^{-}}{\Delta
\omega}
\hat B_t^{-}
\hspace{26mm}
(\hat B_t^{+}
\wedge \hat z
\cdot \hat \epsilon_{ex})(\hat B_t^{-} \cdot \hat \epsilon_{ex}) + ( - \longleftrightarrow  +)  
$ 

\end{table}

 We can also conclude from this discussion how important it is to obtain as much information as possible about the stray
fields and, even more so, to minimize them. We can determine the average value of the stray fields seen by the atoms simply
by exploiting the physical effects analyzed in this work. By applying ``large'' ($\sim$2 G), controlled magnetic fields whose
direction can be chosen and reversed at will, we can isolate the magnetoelectric dichroism described in section 4.1. From its
magnitude, proportional to the stray $\vec E_t$ field and to the applied $\vec B_t$ field, we deduce the averaged 
magnitude and direction of the stray transverse electric field seen by the atoms inside the interaction region. We have
found that the
$\vec E_l $-even contribution is created by a density of electrons circulating inside the cell \cite{gue02}. The field is 
radially distributed around the axis of the cylindrical cell. Since it cancels for optimal centring of the interaction region, we
 can reduce it. In addition, we can correct the $\vec E_l $-odd contribution by making tiny tilts 
(a few $\sim 10^{-3}$ rad) of the cell axis with respect to the common direction of the beams.    
In a similar way we learn about the transverse stray magnetic fields by deliberately amplifying the $\hat B_t^{+} \hat
B_t^{-}$ effects (class~2 effects,\S~5.2), using an applied field $ \vec B_t$ of controlled direction. In order to
disentangle the effects of classes 1 and 2 induced by the applied $\vec B_t$ field, we exploit their different behavior under 
global rotations of the experiment in addition to their different relative size when a different detection probe transition is
used. 

We note finally that optical rotation data are affected by the same stray fields which alter linear dichroism. We use this to
obtain complementary information to monitor and reduce them.

\subsection{Statistical analysis of the data }
A complete set of data is obtained after cycling over four orientations of the input polarizations and provides us with one set of
values of the four quantities
$G_y, G_Y, G_x, G_X$, from which we extract
${\rm S_{xy}}, {\rm S_{XY}}$, the isotropic contributions and ${\rm D}_{_{xy}}, {\rm D}_{_{XY}}$, the signal differences
which may reveal anisotropic contributions (see Eqs.(\ref{sum}) - (\ref{diff})). According to the previous considerations we
can write:
 $${\rm S_{xy}} = {\theta}^{pv} + {\rm S_{xy}^{^{+}}} +  {\rm  S_{xy}^{^{-}}} \,, $$
where we recognize, besides the expected PV contribution, two effects of class 1 corresponding to $\vec E_t,\; \vec B_t$
fields of odd or even behaviour in $\vec E_l$ reversal (superscript + or - respectively). A similar equation holds also for
${\rm S_{XY}}$. On another hand, the difference signal 
$${\rm D_{xy}} = {\rm D_{xy}^{^{+}}}  +  {\rm D_{xy}^{^{-}}} +{\rm  D_{xy}^{^{(2)}}}\;$$
also displays the contribution  ${\rm D_{xy}^{^{(2)}}}$, resulting
from effects of class 2.  It is interesting to underline the connection existing between ${\rm
S_{xy}}$ and ${\rm D_{xy}}$ by reexpressing ${\rm D_{xy}}$ as:
\begin{eqnarray}
{\rm D_{xy}} &  =  & - {\rm  S_{xy}^{^{+}}} \left (\cos{2 \theta_{E_t^{+}}} - \cot{(\theta_{B_t^{+}} - \theta_{E_t^{+}})} \sin{2
\theta_{E_t^{+}}}\right ) + (+ \leftrightarrow - ) 
  + {\rm D_{xy}^{^{(2)}}}\;.\nonumber \\
& = & b^{+} {\rm S_{xy}^{^{+}}} + b^{-} {\rm S_{xy}^{^{-}}} + {\rm D_{xy}^{^{(2)}}} \;.  
\end{eqnarray}
   
 The experimental 
data are to some extent noisy (mainly because of photon noise) and they constitute random variables.  When a
set of data is analyzed there are two questions to be answered: i) is the anisotropy statistically significant? ii) if so, is it
possible to say whether there is an  associated isotropic contribution?    

i) First, in a cartesian coordinate system $({\rm D_{_{x y}}}, {\rm D_{_{X Y}}})$ we plot one point per data set.  Figure 2 illustrates
a cloud of  760 points. We look for any possible deviation of their gravity center with respect to the origin. For the data
points of Fig. 2 this deviation is too small to be significant, it  does not exceed one standard deviation. We conclude 
that this data set shows no significant anisotropy.

ii) Second we evaluate the correlation coefficient r, between
${\rm S_{_{xy}}}$ and ${\rm D_{_{xy}}}$ (and similarly between ${\rm S_{_{XY}}}$ and ${\rm D_{_{XY}}}$).
\begin{equation}
r = \frac {\left ( \sum_{i=1}^{n} {\rm S_{_{xy}}}^{(i)} {\rm D_{_{xy}}}^{(i)} \right )/ n\;  -  <{\rm S_{_{xy}}}>  <{\rm
D_{_{xy}}}>}{s_{_{\rm S}}
\; s_{_{\rm D}}}\,,
\end{equation}
where $<{\rm S_{_{xy}}}>$ and $<{\rm D_{_{xy}}}> $ are the average values and $s_{_{\rm S}}$ and $s_{_{\rm D}}$ the
standard deviations of
${\rm S_{_{xy}}}$ and ${\rm D_{_{xy}}}$ taken over the sample population.   Thus, we
can test the hypothesis that the population correlation is 0 against the alternative that it is not, with a chosen confidence
level. According to Eq. (53), the existence of a linear relation between ${\rm D_{xy}}$ and ${\rm S_{xy}}$ indicates that,
unless the class 2 effects are largely dominant or the class 1 effects absent, we should find a correlation between those two
variables. According to a classical method of statistical analysis, see for instance 
\cite{chao}, if the correlation turns out to be significant, we can obtain from the value of $r$ an estimation of the slope of
the line of regression of ${\rm D_{_{xy}}}$ on ${\rm S_{_{xy}}}$, : $b= \frac {s_{_{\rm D}}}{s_{_{\rm S}}} r$, interpreted as 
$b^{+}$ or
$b^{-}$ (see Eq. (53)).
\begin{figure}
\centerline{\epsfxsize=150mm \epsfbox{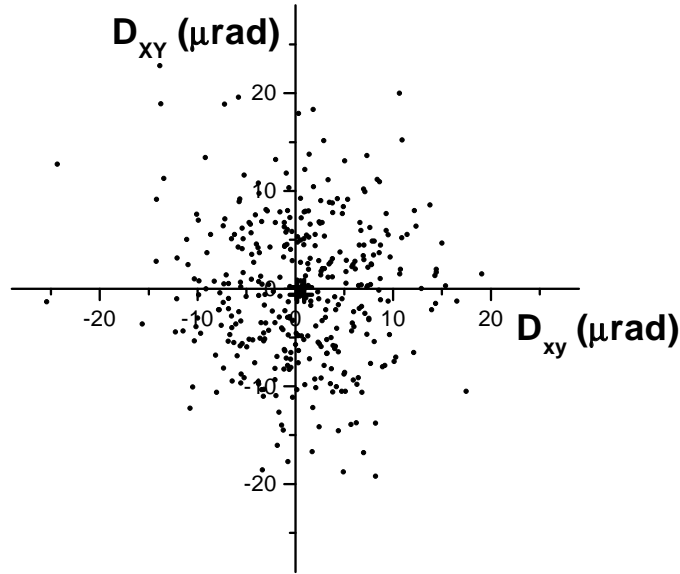}}
\vspace{-15mm}
\caption {\footnotesize  Anisotropy test performed on a sample of 760 sets of plane dichroism data (Eq.
\ref{dichro} ) measured for the four different orientations of the input $\hat \epsilon_{ex} \,, \hat \epsilon_{pr}$
polarizations: for each individual set one  signal difference, $D_{_{XY}} = G_X - G_Y$, is plotted versus  the other one
$D_{_{xy}} = G_x -G_y$. On the cloud of points thus obtained, one looks for a distorsion with respect to a circular
distribution centred on the origin. For the data presented here the center of gravity is indicated. Within the error bars,
$\sigma_D = 1.8 \times 10^{-7}$ rad,  its coordinates merge into the origin.}
\end{figure}

In conclusion, the absence of correlation between the variables ${\rm D_{xy}}$ and ${\rm S_{xy}}$ (and between ${\rm
D_{XY}}$ and
${\rm S_{XY}}$) is an important test. It enables us to conclude whether a non-zero average anisotropy is accompanied by an
isotropic  contribution which might alter the PV signal
we want to detect. More precisely, the fraction of ${\rm D_{xy}}$ which is correlated to ${\rm S_{xy}}$, {\it i.e.} $b \,  {\rm
S_{xy}}$, yields an estimate of the systematic uncertainty which affects $<{\rm S_{xy}}>$. 
\subsection{Order of magnitude of the residual systematic effect}
From the discussion presented in \S~4. and \S~5. it appears that the most worrying  symmetry breaking effects come from
the  isotropic contribution of two stray fields $\vec E_t, \vec B_t$ or of a pump-probe
misalignment $\delta \alpha$ combined with an $\vec E_t ^{+} $ field. An estimate of the resulting systematic effect is
directly derived from Eq. (\ref{EBimb})   
$$ G_{_{EB}} = \, (\hat z \cdot \hat E_l )\, \frac{E_t}{E_l} \omega_{_{F'}} \tau,    $$ where $\omega_{_{F'}} \tau$ is the average
Larmor precession angle of the angular  momentum $\vec F'$ during the time spent in the excited state, and from Eq.
(\ref{misalDat}) 
$$ G_{_{E\delta \alpha}} = \, (\hat z \cdot \hat E_l ) \, \frac{E_t}{E_l} \delta \alpha.  $$ 

 We record and average the harmful defects throughout the whole data
acquisition  by performing auxiliary measurements, at regular time intervals. If, over the whole averaging period, one
achieves  $<{E_t /E_l}> \; \leq 0.5 \times 10^{-3}$ and $<{B_t}>\; \leq 1 $mG  leading to 
$<{\omega_{_{F'}}\tau}> \; \leq 18 \; \mu$rad, we arrive at
$<{G_{_{EB}}}> \; \leq 0.9 \times 10^{-2} \times \theta^{pv}$, and it would make sense to go for one per cent
statistically accurate measurements. Although keeping the defects reduced at this level is difficult to achieve, it does not
look unfeasible. As to the geometrical pump-probe alignment, it is achieved using a reference four-quadrant photo-cell 
sensitive to both the pump and the probe beams  (with respective radius 1.0 and 0.6 mm); this cell is placed alternatively at
the input and at the output of the oven, 30 cm apart. To ensure
$\delta \alpha \leq 18 \times  10^{-6}
$ rad requires a centering quality of  6
$\mu$m on each quadrant cell. This seems achievable, since once achieved, the initial
alignment is preserved during data averaging using servo-loops stabilizing both the pump and the probe
beam positions on the initial reference, making use of auxiliary four-quadrant cells on pick up beams.

Finally, we mention a possible source of systematics which can be generated by the reflection of the pump
beam on the output window of the cell: the reflected beam then goes back through the interaction region with a wave vector badly
aligned with the probe wave vector. If a stray transverse, $\vec E_l$-even electric field is also present, this  favors the systematic
effect $G_{E\delta \alpha}$. This has prompted us to extinguish the reflexion on our cell windows by using temperature
tuning of interferences taking place between the beams reflected by the inner and  outer surfaces of each window \cite{jah00}.
 
\section{Conclusion}
 To exploit the recent progress made using  a new scheme for APV detection in cesium 
\cite{bou02}, we must suppress the systematic effects  in order to match 
 the statistical noise reduction expected  in the next stage of our experiment    
aiming at a one percent precision.  In the present paper,
we  have given a detailed discussion  of a large class of
systematic effects which all break the cylindrical symmetry   
of the ideal experimental  configuration  involved in the excitation and detection processes. 
 A perfect experimental set-up would be invariant under both  symmetries. The first is the mirror reflexion with respect
to the plane defined by the linear polarization of the excitation beam and the electric field 
$\vec E_l$, colinear to this beam; a breaking of this  is  evidence for APV. The second is the cylindrical symmetry around the
common  direction of the pump and probe beams. In an ideal design, the PV signal
would be   invariant under any global rotation of the experiment around  
this direction.

 Experimental defects which break mirror
symmetry do not necessarily break the cylindrical  invariance:  this is 
the case with a longitudinal magnetic field, which happens to be odd under $\vec
E_l$ reversal. The effect of such a field has been considered previously 
\cite{gue02} and we merely recall here that an auxiliary Faraday effect 
measured on a different probe line allows us to correct for it.

 In the present work, our aim was to find  solutions to the broader problem
arising from the experimental defects, namely stray fields and 
misalignments which break both mirror and cylindrical symmetry 
properties. We began by  a short review of the properties of the
chiral optical gain, which gives rise to the PV  signal in our experiment.
This originates from  a mirror symmetry breaking  contribution to   
the excited state atomic alignment, which  reflects  the presence of a PV transition 
amplitude in the excitation process, resulting from the weak PV electron-nucleus interaction.

 On this occasion, we have found it convenient to introduce 
a tensor formalism for calculating the atomic alignment in the excited
state and the detection signal . 
This allows us to incorporate in a systematic way the contributions arising from all 
the 6S-7S transition amplitudes, in presence of transverse $\vec E_t$ and
$\vec B_t$ fields breaking the cylindrical symmetry of the 
experiment. Our two-channel polarimetry measurements performed in
balanced mode provide us with a differential, pseudoscalar signal 
which makes possible dark field detection of the PV effect in
the ideal configuration. We have derived a general expression for 
the differential polarimeter imbalance, this results from the 
contraction of two second-rank tensors, the
alignment tensor and a detection tensor constructed from the 
eigenpolarizations of the polarimeter. 

It is remarkable that both the APV  
and the calibration signals are invariant under simultaneous 
rotations of the input pump and probe polarizations around the
direction common to both beams. During such rotations, stray 
transverse  fields and misalignment remain fixed so that  their relative 
direction
with respect to the beam polarizations are modified. As a result they 
generate new contributions to the atomic polarimeter
imbalance. We have endeavored to answer two crucial questions:  
\begin{itemize} 
\item are these  contributions 
 distinguishable  from the true  PV signal?
\item can we extract enough information from their variations observed
during the rotation of the experiment  to separate  the true  PV signal  
 from systematic effects?
\end{itemize}
 Our analysis  shows that associated with each defect there is either 
a new contribution to the atomic alignment tensor in the
excited state or a tiny rotation of this tensor or (else) of the 
detection tensor.  The tensor formalism introduced at the outset of the paper  is well adapted to derive 
all the corrections to the imbalance to second order in the defects. In order for there to be an  
$\vec E_l$-odd imbalance which simulates the PV signal at least two defects must conspire. The possible 
pairs can be arranged into two classes. Both have their magnitude
characterized by a
$\cos{(2\phi)}$ modulation as the rotations are performed through 180¨$^{\circ}$. After we average  
the results over four configurations obtained by successive
rotations of 45$^\circ$, a systematic effect may only result from the 
effects belonging to the first class, which involves the presence of a 
transverse electric field. To reduce the  average transverse $\vec E_t$
field seen by the atoms, we perform auxiliary sequences of measurements with a known 
applied magnetic field. We must also compensate carefully the transverse
magnetic fields, using the atomic 
signals described in this work.

After  optimization, there is the possibility  of a left-over anisotropic contribution which 
could be associated both with class~1  and class~2 processes. We have found 
a way to get information about the harmful   
 class 1 isotropic contribution  by  showing  that 
its fluctuations are correlated  
with those of the class~1 anisotropy contribution.
( No such correlation of course can exist for the APV purely isotropic term). A statistical treatment of the   
data yields an estimate of the systematic uncertainty 
associated with the class~1 processes.

As an aside  remark, we would like to point out that,  
during the course of this work, we have established a 
connection between the most worrying sources of systematics,
generated by parallel components of transverse electric and magnetic 
fields, and the magnetoelectric Jones dichroism \cite{jon48}
requiring  extreme conditions for being observed in liquid samples 
\cite{rik01}. This highlights  the extreme sensitivity of  highly forbidden transtions,
such as the cesium $6S-7S$, to the symmetry of the experimental set-up 
and illustrates the great variety of new processes which can be studied both
theoretically and experimentally.

\section*{Acknowledgements}
We are very grateful to Claude Bouchiat for many fruitul discussions and to Mark Plimmer 
for critical reading of the manuscript.

 \end{document}